\documentclass[preprint,aip,jcp]{revtex4-1}
\usepackage[utf8]{inputenc}
\usepackage[version=3]{mhchem}
\usepackage{titlecaps}
\usepackage{chemformula} 
\usepackage[T1]{fontenc} 
\usepackage{enumitem}
\usepackage{graphicx}
\usepackage{bm}

\Addlcwords{a an the of on at with for and in into from by without}

\begin{document}

\author{Julian Arnold} \affiliation{Department of Physics, University
  of Basel, Klingelbergstrasse 80, CH-4056 Basel, Switzerland}
\affiliation{Department of Chemistry, University of Basel,
  Klingelbergstrasse 80, CH-4056 Basel, Switzerland}

\author{Juan Carlos San Vicente Veliz} \affiliation{Department of
  Chemistry, University of Basel, Klingelbergstrasse 80, CH-4056
  Basel, Switzerland}

\author{Debasish Koner}\affiliation{Department of Chemistry,
  University of Basel, Klingelbergstrasse 80, CH-4056 Basel,
  Switzerland}\affiliation{Department of Chemistry, Indian
  Institute of Science Education and Research (IISER) Tirupati,
  Karakambadi Road, Mangalam, Tirupati 517507, Andhra Pradesh, India}

\author{Narendra Singh} \affiliation{ Department of Mechanical
  Engineering, Stanford University , Stanford, CA 94305, USA}

\author{Raymond J. Bemish} \affiliation{Air Force Research Laboratory,
  Space Vehicles Directorate, Kirtland AFB, New Mexico 87117, USA}

\author{Markus Meuwly}\email{m.meuwly@unibas.ch}
\affiliation{Department of Chemistry, University of Basel,
  Klingelbergstrasse 80, CH-4056 Basel, Switzerland}

\date{\today}

\title[]{Machine Learning Product State
  Distributions from Initial Reactant States for a Reactive
  Atom-Diatom Collision System}

\begin{abstract}
A machine learned (ML) model for predicting product state
distributions from specific initial states (state-to-distribution or
STD) for reactive atom-diatom collisions is presented and
quantitatively tested for the N($^4$S)+O$_{2}$(X$^3 \Sigma_{\rm
  g}^{-}$) $\rightarrow$ NO(X$^2\Pi$) +O($^3$P) reaction. The
reference data set for training the neural network (NN) consists of
final state distributions determined from explicit quasi-classical
trajectory (QCT) simulations for $\sim 2000$ initial
conditions. Overall, the prediction accuracy as quantified by the
root-mean-squared difference $(\sim 0.003)$ and the $R^2$ $(\sim
0.99)$ between the reference QCT and predictions of the STD model is
high for the test set and off-grid state specific initial conditions
and for initial conditions drawn from reactant state distributions
characterized by translational, rotational and vibrational
temperatures. Compared with a more coarse grained
distribution-to-distribution (DTD) model evaluated on the same initial
state distributions, the STD model shows comparable performance with
the additional benefit of the state resolution in the reactant
preparation. Starting from specific initial states also leads to a
more diverse range of final state distributions which requires a more
expressive neural network to be used compared with DTD. Direct
comparison between explicit QCT simulations, the STD model, and the
widely used Larsen-Borgnakke (LB) model shows that the STD model is
quantitative whereas the LB model is qualitative at best for
rotational distributions $P(j')$ and fails for vibrational
distributions $P(v')$. As such the STD model can be well-suited for
simulating nonequilibrium high-speed flows, e.g., using the direct
simulation Monte Carlo method.
\end{abstract}

\maketitle

\section{Introduction}
Predicting the outcomes of chemical reactions is one of the essential
tasks for efficient material design, engineering, or reaction
planning.\cite{MM.rev:2021} Understanding chemical reactions at a
molecular level can also shed light on the mechanisms underlying
chemical transformations. However, the exhaustive characterization of
reactions at the microscopic (i.e. state-to-state or STS) level
quickly becomes computationally intractable using conventional
approaches due to the rapid growth of the underlying state
space.\cite{grover:2019,MM.nncs:2019} As an example, even for a
reactive atom+diatom system (A+BC$\rightarrow$AB+C) the number of
internal states for diatoms AB and BC is $\sim 10^4$ which leads to
$\sim10^8$ state-to-state cross sections $\sigma_{v,j \rightarrow
  v',j'}(E_{\rm trans})$ between initial $(v,j)$ and final $(v',j')$
rovibrational states at a given relative translational energy $E_{\rm
  trans}$.\cite{MM.nncs:2019} The estimated number of classical
trajectories required for converged STS cross sections is $\sim
\mathcal{O}(10^{13})$ assuming that $10^5$ classical trajectories are
sufficient for one converged cross section. For reactive diatom+diatom
systems this number increases to $\sim \mathcal{O}(10^{20})$ which is
currently unfeasible.\cite{schwartzentruber:2018}\\

\noindent
Machine learning (ML) methods are well suited for such tasks as they
are designed for large data sets and generalize well towards unseen
input data.\cite{goodfellow:2016,MM.rev:2021} In particular, neural
network (NN)-based models have successfully been used to predict the
STS cross sections of reactive atom-diatom collision
systems.\cite{MM.nncs:2019} These models were trained on data obtained
from explicit quasi-classical trajectory (QCT) simulations. Similarly,
NN-based models were constructed at the distribution-to-distribution
(DTD) level.\cite{MM.dtd:2020} For a given set of distributions of
initial states of reactants ($P(E_{\rm trans})$, $P(v)$, $P(j)$), a
DTD model aims at predicting the relative translational energy
distribution $P(E_{\text{trans}}')$, together with the vibrational
$P(v')$ and rotational $P(j')$ state distributions of the
product. Compared with a STS model, state specificity is lost in a DTD
model as it follows how a distribution of initial reactant states is
processed through interactions on a potential energy surface (PES),
but does not keep track of the interrelations between individual
initial and final states. This information loss makes DTD models
computationally cheaper compared with STS models.\\

\noindent
Motivated by these findings, the present work explores the possibility
to conceive an intermediate model between the STS and DTD models which
retains state specific information for the reactants. In the following
it is demonstrated how a NN-based state-to-distribution (STD) model
for a reactive atom+diatom system can be developed. The STD model is
shown to predict product state distributions $P(E^{'}_{\rm trans}),
P(v')$, and $P(j')$ given a specific initial reactant state $(E_{\rm
  trans}, v, j)$. The necessary reference data to train such a
NN-based STD model was obtained from explicit quasi-classical
trajectory (QCT) simulations for the N($^4$S)+O$_{2}$(X$^3 \Sigma_{\rm
  g}^{-}$) $\rightarrow$ NO(X$^2\Pi$) +O($^3$P) collision system as a
proxy. As such, an STD model may be constructed from a STS model
through coarse graining, i.e. by integration of the final
states. Similarly, a DTD model can be obtained from an STD model by
further coarse graining of the state-specific initial conditions. Note
that such a coarse graining by means of integration does, however,
incur a computational overhead. Moreover, the increase in information
content going from a DTD model to a STD model, and finally to a STS
model, also comes at an increased number of trainable parameters and,
hence, increased computational cost both in training and evaluation of
the model. Therefore, it is crucial to choose the appropriate model
resolution for a given task. Finally, it is shown that the STD model
realizes a favourable trade-off between computational cost and
accuracy, i.e., information content. In particular, the STD model
provides information at an appropriate resolution to be utilized as
input for methods such as Direct Simulation Monte Carlo
(DSMC)\cite{dsmc:2017} or computational fluid dynamics (CFD)
simulations.\cite{walpot:2012}\\

\noindent
This work is structured as follows. First, the methods including the
data generation based on quasi-classical trajectory simulations, as
well as the neural network architecture and its training are
described. Next, the ability of the STD model to predict product state
distributions from unseen, specific initial states of the reactant is
assessed. Then, the differences between DTD and STD models at
predicting product state distributions from initial state
distributions is discussed. Finally, the performance of the STD model
is compared with the widely used
Larsen-Borgnakke~\cite{borgnakke1975statistical} for simulations of
nonequilibrium, high-speed flows, and then conclusions are drawn.

\section{Methods}
\subsection{Quasi-Classical Trajectory Simulations}
\label{QCT}
Explicit QCT calculations for the N + O$_{2} \rightarrow$ NO + O
reaction were carried out following previous
work.\cite{tru79,hen11,kon16:4731,Koner2018,MM.no2:2020} Specifically,
the reactive channel for NO formation (N($^4$S)+O$_{2}$(X$^3
\Sigma_{\rm g}^{-}$) $\rightarrow$ NO(X$^2\Pi$) +O($^3$P)) was
considered here. For this, the $^4$A$'$ PES was chosen as NO formation
is dominated by contributions from the $^4$A$'$ electronic
state.\cite{MM.no2:2020} Hamilton's equations of motion were solved in
reactant Jacobi coordinates using a fourth-order Runge-Kutta method
with a time step of $\Delta t = 0.05$ fs, which guarantees
conservation of the total energy and angular
momentum.\cite{kar65:3259,Koner2018}\\

\noindent
For generating the training, test, and validation data set for the NN
the following {\it state-specific} initial conditions were used: $(0.5
\leq E_{\rm trans} \leq 8.0)$ eV with $\Delta E_{\rm trans}=0.5$ eV;
$v=[0,2,4,6,8,10,12,15,18,21,24,27,30,34,38]$; and $0 \leq j \leq 225$
with $\Delta j = 15$, resulting in 2184 different states. The impact
parameter $b$ was sampled from 0 to $b_{\rm max}=12$ a$_{0}$ using
stratified sampling.\cite{tru79,bender:2015} Ro-vibrational states of
the reactant (O$_{2}$) and product diatom (NO) are determined from the
semiclassical theory of bound states.\cite{kar65:3259} First, final
vibrational and rotational states were determined as real numbers from
the diatomic internal energy and angular momentum, respectively,
whereas the translational energy is obtained from the relative
velocity of the atom+diatom system. Ro-vibrational quantum numbers are
then assigned as the nearest integers $(v',j')$ using the histogram
binning method. To conserve total energy, the ro-vibrational energy
$E_{v' j'}$ is recomputed from semiclassical
quantization\cite{kar65:3259,tru79} using the quantum numbers
$(v',j')$ and the final translational energy for the atom+diatom
system is adjusted using $E_{\rm trans}^{'} = E_{\rm tot}
-E_{v' j'}$. Product states were assigned using histogram binning ($0.1
\leq E_{\rm trans}' \leq 19.8$) eV; $0 \leq v' \leq 47$ with $\Delta
v' = 1$; $0 \leq j' \leq 240$ with $\Delta j' = 1$. Out of the 2184
initial reactant states, 7 (with $E_{\rm trans}=0.5$ eV) resulted in
product state distributions with zero or negligible probability
(max($P) < 10^{-5}$) which were not considered for the subsequent
analysis. Consequently, 2177 initial reactant states together with the
corresponding product state distributions obtained by QCT simulations
constitute the reference data to train and test NN-based STD models in
this work. \\

\noindent
To evaluate the trained models, a second set of initial conditions was
generated from reactant {\it state distributions}. For each trajectory
they were randomly chosen using standard Monte Carlo
methods.\cite{tru79,hen11} The initial relative translational energies
$E_{\rm trans}$ were sampled from Maxwell-Boltzmann distributions
$(0.0 \leq E_{\rm trans} \leq 19.8)$ eV with $\Delta E_{\rm trans} =
0.1$ eV. Vibrational $(v)$ and rotational $(j)$ states were sampled
from Boltzmann distributions, where $0 \leq v \leq 38$ with $\Delta v
= 1$; and $0 \leq j \leq 242$ with $\Delta j = 1$. These distributions
are characterized by $T_{\text{trans}}$, $T_{\text{vib}}$, and
$T_{\text{rot}}$, respectively.\cite{tru79,bender:2015} For each set
of temperatures $\bm{T}= (T_{\text{trans}}$, $T_{\text{vib}}$,
$T_{\text{rot}}$), 80000 trajectories were run to obtain the product
state distributions. First, models were constructed with
$T_{\text{rovib}}=T_{\text{rot}}=T_{\text{vib}}$, for which QCT
simulations were performed at $T_{\text{trans}}$ and
$T_{\text{rovib}}$ ranging from 5000 K to 20000 K in increments of 250
K. This yielded 3698 sets of reactant states and corresponding product
state distributions. Next, for the more general case $T_{\text{rot}}
\neq T_{\text{vib}}$, additional QCT simulations were performed for
$T_{\text{trans}} = 5000, 10000, 15000, 20000$ K with $T_{\text{vib}}$
and $T_{\text{rot}}$ each ranging from 5000 K to 20000 K in increments
of 1000 K. Combining these additional 960 data sets with the 3698 sets
from above leads to a total of 4658 data sets.\\

\subsection{Data Preparation}
An important step in conceiving a ML model is the preparation,
representation and featurization of the data. For featurization the
following properties were chosen as input to the NN: 1.)
$E_{\text{trans}}$, 2.)  vibrational quantum number $v$ of the
diatomic, 3.) rotational quantum number $j$ of the diatomic, 4.)
relative velocity of diatom and atom, 5.) internal energy $E_{v,j}$,
of the diatom, 6.)  vibrational energy $E_{v,j=0}$, of the diatom, 7.)
rotational energy $E_{v=0,j}$, of the diatom, 8.)  angular momentum of
the diatom, 9-10.) the two turning points at each of the vibrational
states of the reactant diatom, and 11.)  the vibrational time period
of the diatom. These features were already used successfully for the
STS model.\cite{MM.nncs:2019}\\

\begin{figure}[b!]
\begin{center}
\includegraphics[width=0.99\textwidth]{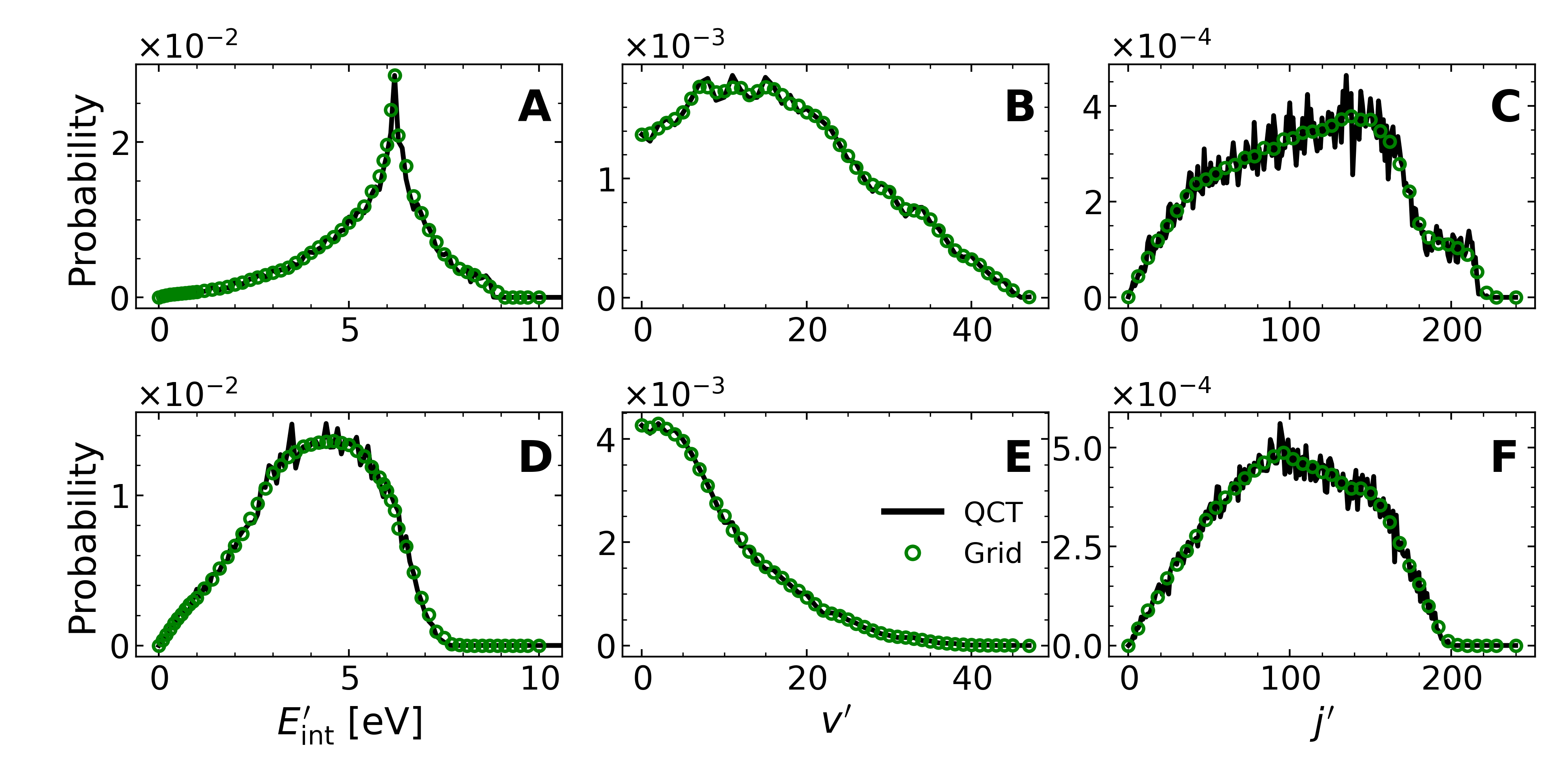}
\caption{Product state distributions $P(E_{\rm int}^{'})$, $P(v')$,
  and $P(j')$ obtained from QCT simulations (QCT), as well as the
  corresponding amplitudes that serve as a reference for training and
  testing the NN-based STD models (Grid). The product state
  distributions correspond to initial reactant states characterized
  by: (A to C) ($E_{\text{trans}}=3.0$ eV, $v=34$, $j=0$;
  $E_{\text{int}}=4.8$ eV), (D to F) ($E_{\text{trans}}=5.0$ eV,
  $v=6$, $j=45$; $E_{\text{int}}=1.5$ eV).}
\label{fig:fig2}
\end{center}
\end{figure}

\noindent
To represent the product state distributions, a grid-based (G-based)
approach was used.\cite{MM.dtd:2020} In a G-based approach, each
product state distribution is characterized by its values at discrete
grid points, referred to as ``amplitudes'' in the following. Figure
\ref{fig:fig2} shows the product state distributions from QCT
simulations (solid line) and their G-based representation (open
symbols) for two exemplar reactant states. The G-based representation
closely follows the true, underlying data from the QCT
simulations. Thus, G-based product state distributions
(i.e. amplitudes) are suitable to train the NN and the amplitudes also
constitute the output of the trained NN. Calculating the product state
amplitudes for all available data sets then allowed to train and test
the NN. Subsequently, inter- and extrapolation can be performed to
obtain a continuous prediction. This is referred to as the ``STD
model'' in the following.\\

\begin{figure}[t!]
\begin{center}
\includegraphics[width=0.99\textwidth]{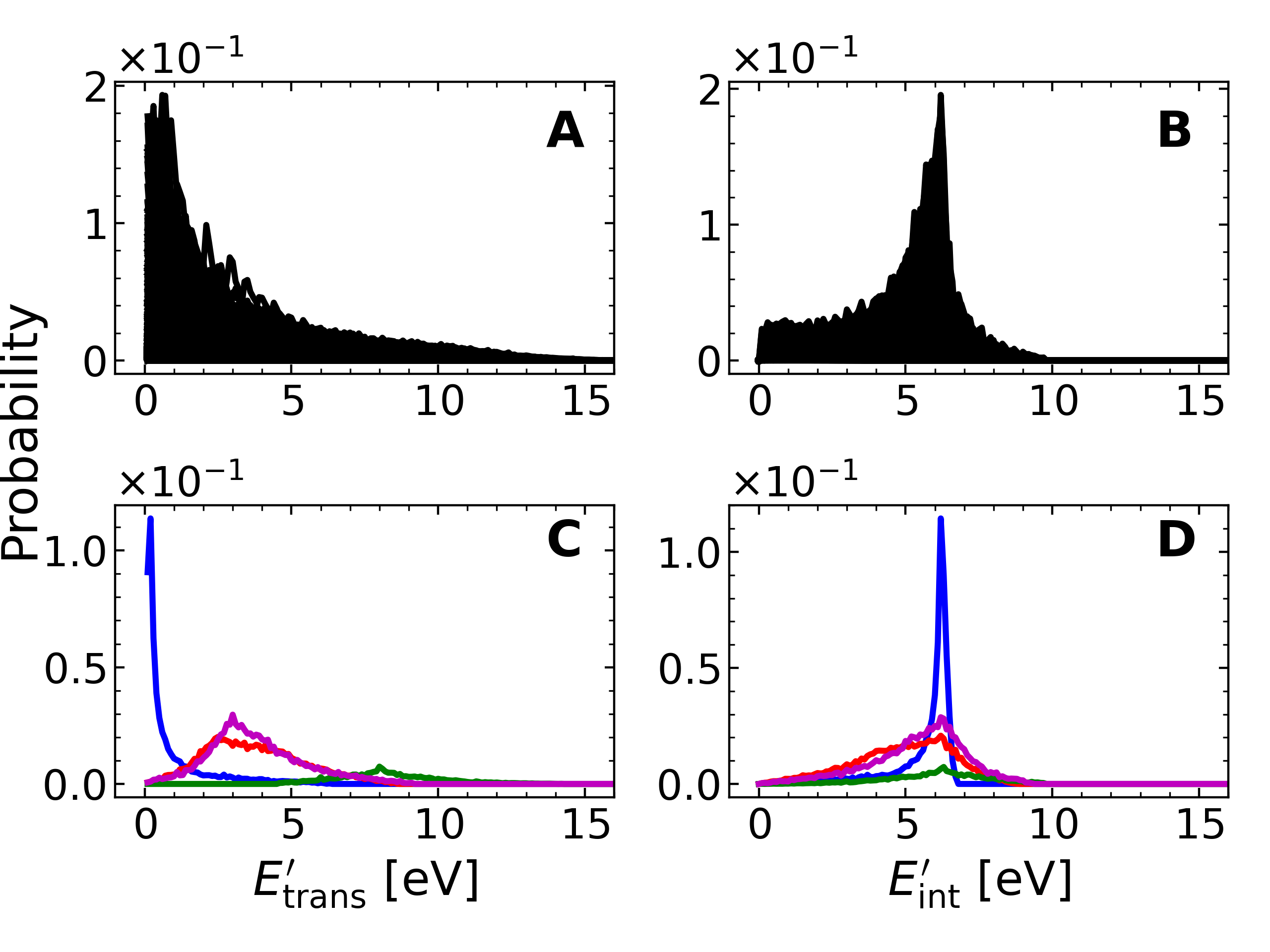}
\caption{Distributions $P(E_{\rm trans}')$ (panel A) and $P(E_{\rm
    int}^{'})$ (panel B) obtained from QCT simulations for each of the
  2184 initial reactant states considered in this work displayed on
  top of each other. Panels C and D: a few selected distributions
  $P(E_{\rm trans}')$ and $P(E_{\rm int}^{'})$ to highlight their
  different shapes that motivate the choice of $P(E_{\rm int}^{'})$
  over $P(E_{\rm trans}')$.}
\label{fig:pofe}
\end{center}
\end{figure}

\noindent
For the product state distributions it was found to be advantageous to
consider the set ($P(E_{\rm int}^{'}),P(v'),P(j')$) instead of
($P(E_{\text{trans}}'),P(v'),P(j')$). Here, $E_{\rm int}^{'} = E_{\rm
  tot} - E_{\rm trans}'$ is the internal energy after removing the
translational energy. Note that $P(E_{\rm int}^{'})$ and
$P(E_{\text{trans}}^{'})$ contain the same information and can be
interconverted because the total energy $E_{\rm tot}$ of the system is
conserved. However, for representing $P(E_{\rm int}^{'})$ fewer grid
points are required than for representing $P(E_{\text{trans}}')$. This
is illustrated in Figure \ref{fig:pofe}, where $P(E_{\text{trans}}')$
and $P(E_{\rm int}^{'})$ obtained from explicit QCT simulations for
all 2184 initial reactant states used to train and validate in this
work are shown. While there are distributions $P(E_{\text{trans}}')$
which are non-zero at $E_{\text{trans}}'> 10$ eV, all $P(E_{\rm
  int}^{'})$ are zero for $E_{\rm int}^{'} > 10$ eV and grid points
are only used up to this value.\\

\noindent
The location and number of grid points to represent the product state
distributions was motivated after inspection of the overall shape of
these distributions. In particular, it was observed that a large
number of $P(E_{\rm int}^{'})$ distributions exhibit a sharp peak for
$E_{\rm int}^{'}\sim 6.2$ eV which is the dissociation energy of the
product diatom NO (see Figures \ref{fig:fig2}A, \ref{fig:pofe}B and
D).\cite{luo:2012} Non-zero contributions to $P(E_{\rm int}^{'})$ at
$E_{\rm int}^{'}$ larger than the dissociation energy of the product
diatom NO can be attributed to the presence of quasi-bound
states. Additionally, $P(E_{\rm int}^{'})$ can also increase rapidly
for $E_{\rm int}^{'} < 1.0$ eV. Consequently, the grid for $P(E_{\rm
  int}^{'})$ was chosen more densely for $E_{\rm int}^{'} < 1.0$ eV
and $E_{\rm int}^{'} \sim 6.2$ eV to capture these features ($E_{\rm
  int}^{'} = [0.0, 0.1, \cdots, 1.0], [1.2, 1.4, \cdots , 6.0], [6.1,
  6.2, 6.3], [6.5, 6.7, \cdots , 9.7, 10.0]$ eV).\\

\noindent
A considerable number of the final state vibrational distributions,
$P(v')$, show a maximum for $v' \sim 0$ (see Figures \ref{fig:fig2}B
and \ref{fig:fig2}E). For higher $v'$, $P(v')$ typically decays
rapidly but in general, the distributions display a variety of
shapes. Hence, the corresponding grid was dense ($0 \leq v' \leq 47$
with $\Delta v' = 1$). Final state rotational distributions, $P(j')$,
are closer in overall shape to one another compared with $P(E_{\rm
  int}^{'})$ or $P(v')$. In particular, $P(j')$ typically does not
exhibit sharp features (see Figures \ref{fig:fig2}C and D). Taking
this into consideration, the grid for $P(j')$ was equidistant ($0 \leq
j' \leq 240$ with $\Delta j' = 6$) and less dense than for the other
two final state distributions.\\

\noindent
The number of grid points for ($E_{\rm int}^{'},v',j'$) was $(58, 47,
40)$, respectively. This is significantly more dense than the DTD
model,\cite{MM.dtd:2020} for which $(16,16,12)$ grid points were used
and is attributed to the fact that the distributions considered here
are more diverse and exhibit more detail, including sharp
features. The shapes of the distributions $P(E_{\rm int}')$, $P(v')$
or $P(j')$ are generally smooth across the ranges of $E_{\rm int}'$
and quantum numbers $v'$ and $j'$. They also tend to vary smoothly as
the initial state changes. However, when reaction channels open there
can be sharp features in the probability distribution, see Figure
\ref{sifig:fig4}. Because the grids used here are dense, linear
interpolation can be used to obtain a continuous NN-based prediction
of product state distributions at off-grid points.\\

\noindent
Instead of directly sampling the product state distributions $P(x)$ at
the grid points $x_{i}$ to obtain the amplitudes in the G-based
representation, local averaging according to
\begin{equation}
\bar{P}(x_{i}) = \frac{1}{2n+1}\sum\limits_{j=i-n}^{i+n} P(x_{j}),
\label{eq:local_averaging}
\end{equation}
was performed. Here, the number of neighbouring data points $x_{j}$
(not necessarily grid points) considered for averaging is $n \in [0,
  n_{\rm max}]$. If there are fewer neighbouring data points to the
right and/or to the left of grid point $x_{i}$ when compared with
$n_{\text{max}}$, $n$ was chosen as the maximum number of neighbouring
data points available to both sides, otherwise
$n=n_{\text{max}}$. Consequently, the first and last data points were
assigned unaveraged values. Note, that the value of $n_{\text{max}}$
can differ for each of the 3 degrees of freedom
$(E_{\text{int}}',v',j')$. Additionally, no local averaging was
performed for ``sharp'' peaks and only a reduced amount was applied at
nearby points. A maximum was classified as ``sharp'' if the slopes of
the two lines fit to neighbouring data points to the left and right of
it exceeded a given threshold, see Section~\ref{si_sec_peak_detection}
in the SI for details.\\

\subsection{Neural Network}
\label{sec:nn}
\begin{figure}[h!]
\includegraphics[width=0.35\textwidth]{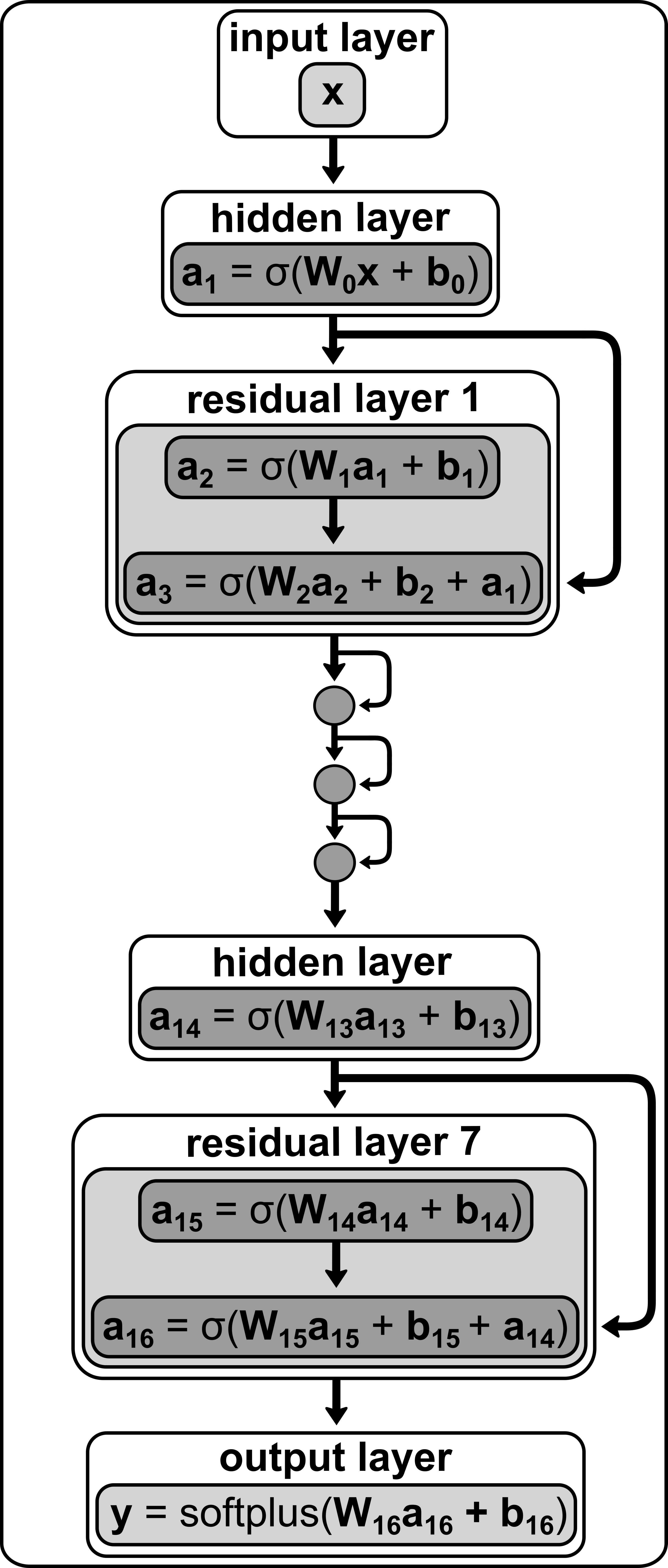}
\caption{Schematic for the NN architecture underlying the STD
  model. The activation vector of each layer is denoted as
  \textbf{a$_{i}$}, and the input and output vectors are $\mathbf{x}$
  and $\mathbf{y}$. The weight matrix and bias vector for each layer
  are denoted by $\mathbf{W}_{i}$ and $\mathbf{b}_{i}$,
  respectively. The activation function of the hidden layers is
  $\sigma(\bm{z})$ and corresponds to a shifted
  softplus\cite{dugas:2001} function $\sigma$($\bm{z}$) =
  $\ln(1+e^{\bm{z}})-\ln(2)$, where
  softplus($\bm{z}$)$=\ln(1+e^{\bm{z}})$.\cite{clevert:2015,lecun:2012}
  Activation functions act element-wise on vectors.}
\label{fig:fig1}
\end{figure}

The NN architecture for the STD model is shown in Figure
\ref{fig:fig1} and is inspired by ResNet.\cite{he:2016} The input and
output layers consist of 11 inputs (the 11 features, see above) and
$58 + 47 + 40 = 145$ output nodes for the amplitudes characterizing
the product state distributions. The main part of the NN consists of 7
residual layers, each of which is again composed of two hidden layers,
and two separate hidden layers. The shortcut connections,
characteristic for residual layers, help to address the vanishing
gradient problem.\cite{he:2016} Hidden layers 1 to 14 are each
composed of 11 nodes, whereas hidden layers 14 to 16 are each composed
of 44 nodes which leads to a ``funnel-like'' NN architecture that
helps to bridge the gap between the small number of inputs (11) and
the large number of outputs ($58 \times 47 \times 40 = 109040$). The
NN for the STD model has 3746 trainable parameters compared with
$\approx 140$ parameters that are used in the DTD
model.\cite{MM.dtd:2020} The larger number of trainable parameters for
STD can be attributed to the larger diversity of product state
distributions requiring a denser grid, i.e., a wider output
layer. Moreover, the decrease in the number of free model parameters
in going from STD to DTD reflects the reduced information content
which is the state-specificity on the reactant side.\\

\noindent
Before training the NN inputs were standardized via the transformation
\begin{equation}
\label{standardization}
x_{i}' = (x_{i} - \bar{x}_{i})/\sigma_{i},
\end{equation}
and the NN outputs are normalized
\begin{equation}
\label{normalization}
x_{i}' = x_{i}/\sigma_{i},
\end{equation}
where $x_{i}$ denotes the $i$-th input/output (as specified above),
and $\bar{x}_{i}$ and $\sigma_{i}$ are the mean and standard deviation
of the distribution of the $i$-th input/output over the entire
training data. Standardization results in distributions of the
transformed inputs $x_{i}'$ over the training data that are
characterized by ($\bar{x}'_{i}=0$, $\sigma'_{i}=1$) and allows
prediction of high- and low-amplitude data with similar
accuracy. Also, standardization generally yields faster convergence of
the gradient-based optimization.\cite{lecun:2012} The distributions of
the transformed outputs $x_{i}'$ over the training data have
($\bar{x}'_{i}=\bar{x}_{i}$, $\sigma'_{i}=1$) through
normalization. This enables the use of a root-mean-squared deviation
(RMSD) loss function
\begin{equation}
  \label{loss_func}
\mathcal{L} = \sqrt{\frac{1}{N}\sum\limits_{i=1}^{N} \left( y_{i} -
  y'_{i}\right)^2}.
\end{equation}
where $y_{i}$ and $y'_{i}$ denote the value of the $i$-th predicted
and reference amplitude, respectively. Unnormalized output may
drastically differ in amplitude and spread which can lead to poor
performance of the RMSD loss. However, this ignores inherent sampling
noise arising from potentially unconverged QCT simulations, a point
considered explicitly in the following. Using a softplus activation
function for the output layer was found to significantly increase the
NN prediction accuracy compared in contrast to a scaled hyperbolic
tangent. Specifically, using softplus removes unphysical undulations
and unphysical negative probabilities which would arise in the
predicted product distributions in regions where the corresponding
reference distributions are small or zero.\\

\noindent
The weights and biases of the NN were initialized according to the
Glorot scheme\cite{glorot2010understanding} and optimized using
Adam\cite{kingma2014adam} with an exponentially decaying learning
rate. The NN was trained using TensorFlow\cite{tf:2016} and the set of
weights and biases resulting in the smallest loss as evaluated on the
validation set were subsequently used for predictions. From the total
number of $N_{\rm tot}=2177$ data sets, $N_{\rm train}=1700$ were
randomly selected for training, $N_{\rm valid}=400$ were used for
validation and $N_{\rm test}=77$ were used as the test
set.\cite{MM.nncs:2019} All NNs underlying the STD models in this work
were trained on a 3.6 GHz Intel Core i7-9700k CPU resulting in
training times shorter than 4 minutes.\\

\section{Results}
First, the performance of the STD model in predicting product state
distributions given specific initial reactant states is
discussed. This is done for the test set ($N_{\rm test}=77$) and a
considerably broader set of initial conditions not covered in training
or validation (off-grid). In a next step, the capability of the STD
model to predict product state distributions given distributions over
initial reactant states is assessed. These include distributions with
$T_{\rm trans} = T_{\rm vib} = T_{\rm rot}$, $T_{\rm trans} \neq
T_{\rm vib} = T_{\rm rot}$, $T_{\rm trans} = T_{\rm rot} \neq T_{\rm
  vib}$ as the most relevant case for hypersonics, and $T_{\rm trans}
\neq T_{\rm vib} \neq T_{\rm rot}$ as the most general case. The
results for $T_{\rm vib} \neq T_{\rm rot}$ are also compared with
those from the DTD model\cite{MM.dtd:2020}.\\

\subsection{Performance for Given Initial States}
The performance measures to assess the quality of the STD model
considered are ${\rm RMSD} = \sqrt{\sum_{i=1}^{N} \frac{(P_{i} -
    O_{i})^{2}}{N}}$ and $R^{2} = 1 - \sum_{i=1}^{N} \frac{(P_{i} -
  O_{i})^{2}}{(O_{i} - \langle O \rangle)^{2}}$ where $P_i$ is the
predicted value $i$ from STD, $O_i$ is the observed (reference) value
$i$ from QCT, and $\langle O \rangle$ is the average for a given
initial condition and a given degree of freedom. The performance
measures are determined for all three degrees of freedom individually
and for their entirety. The subscript ``LG'' refers to evaluating the
STD and QCT models only on the locally averaged grid points, whereas
the subscript ``FG'' refers to using all grid points at which QCT data
is available (full grid). For this comparison the reference and
predicted amplitudes are first normalized with the normalization
calculated by numerical integration of the reference QCT
distributions. Predictions for STD at off-grid points are obtained
through linear interpolation.\\

\begin{table}[ht]
\centering
\begin{tabular}[t]{l|cc|cc}
\hline
\hline
STD model&${\rm RMSD}_{\rm LG}$&${R^2}_{\rm LG}$ & ${\rm RMSD}_{\rm FG}$&${R^2}_{\rm FG}$\\
\hline
overall& 0.0039 & 0.9886  & 0.0033 & 0.9890  \\
$E_{\rm int}^{'}$&0.0095 & 0.9915 & 0.0077 & 0.9906  \\
$v'$& 0.0020 & 0.9885  & 0.0018 & 0.9895  \\
$j'$& 0.0003 & 0.9860  & 0.0003 & 0.9867  \\
\hline
\hline
\end{tabular}
\caption{Performance measures ${\rm RMSD}$ and $R^2$ for the test set
  $(N_{\rm test} = 77)$. The mean error is calculated separately using
  the distributions of $E_{\rm int}^{'}$, $v'$, or $j'$ and then
  averaged to obtain an overall performance measure. Subscripts ``LG''
  and ``FG'' refer to the ``local grid'' (on which STD is evaluated)
  and ``full grid'' (on which the reference QCT results are
  available).}
\label{tab:1}
\end{table}%

\noindent
The performance measures of the STD model on the test set are
summarized in Table~\ref{tab:1}. Overall, $\rm {RMSD}_{\rm LG}=0.0039$
and $R^2_{\rm LG}=0.9886$ values confirm that the NN gives highly
accurate predictions of the amplitudes on a grid characterizing the
product state distributions. The performance is preserved even for the
``full grid'' (FG). The decreasing performance for predicting
$P(E_{\rm int}^{'})$ compared to $P(v')$ or $P(j')$ (see RMSD and
$R^2$ in Table~\ref{tab:1}) can be attributed to the fact that
$P(E_{\rm int}^{'})$ varies strongly in shape but is typically peaked
which is challenging to capture using a G-based approach. In contrast,
$P(j')$ varies least and can thus be predicted with the highest
accuracy as can be seen from the lowest RMSD and $R^2$ values. The
small difference in accuracy when comparing RMSD$_{\rm LG}$ and
$R^{2}_{\rm LG}$ to ${\rm RMSD}_{\rm FG}$ and $R^{2}_{\rm FG}$ arises
because linear interpolation is used to obtain predicted amplitudes
between the designated grid points.\\

\begin{figure}[bth!]
\begin{center}
\includegraphics[width=0.99\textwidth]{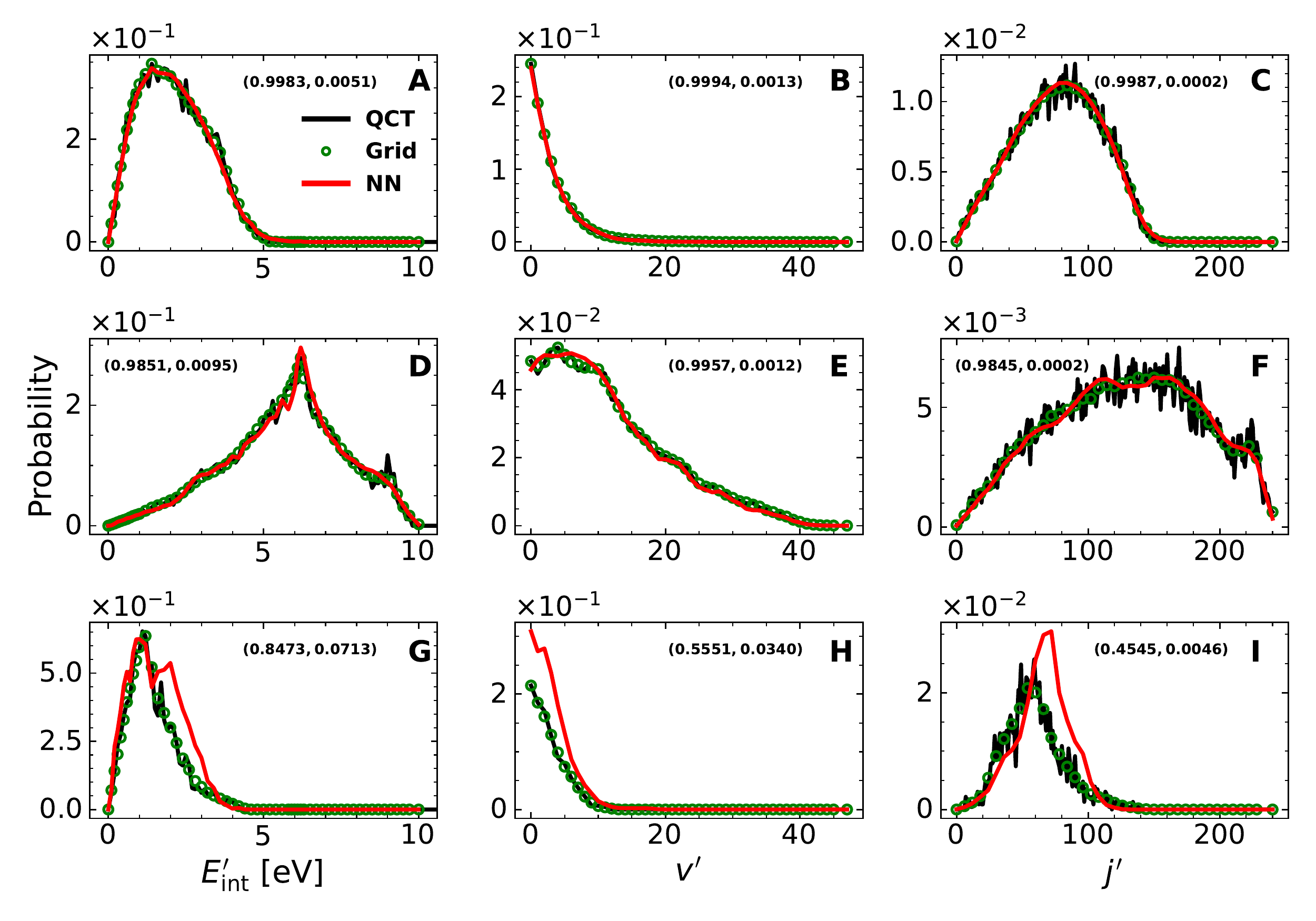}
\caption{Product state distributions obtained from explicit QCT
  simulations (QCT), and the corresponding reference amplitudes (Grid)
  and STD model predictions (NN) for three initial conditions from the
  test set (77 sets) not used in the training. The predictions for
  these three data sets are characterized by (A to C) a $R^{2}_{\rm
    LG}$ value closest to the mean $R^{2}_{\rm LG}$ value as evaluated
  over the entire test set, (D to F) the largest and (G to I) smallest
  $R^{2}_{\rm LG}$ value in the test set, respectively. (A to C)
  ($E_{\text{trans}}=3.5$ eV, $v=0$, $j=45$, $E_{v,j}=0.85$ eV), (D to
  F) ($E_{\text{trans}}=6.0$ eV, $v=21$, $j=0$, $E_{v,j}=1.11$ eV), (G
  to I) ($E_{\text{trans}}=0.5$ eV, $v=0$, $j=135$, $E_{v,j}=0.32$
  eV). For each distribution $(R_{\rm LG}^2, {\rm RMSD_{\rm LG}})$
  values are provided.}
\label{fig:fig5}
\end{center}
\end{figure}

\noindent
Predictions of the STD model for three different sets of initial
reactant states from the test set are shown in
Figure~\ref{fig:fig5}. These data sets are characterized by 1) a
$R_{\rm LG}^2$ value closest to the average $R_{\rm LG}^2$ value over
the entire test set (77 sets) (panels A to C), and $R_{\rm LG}^2$
values corresponding to 2) the largest (panels D to F; ``best
performing'') and 3) the smallest (panels G to I; ``worst
performing'') $R_{\rm LG}^2$ values in the test set, respectively. The
amplitudes of the product state distributions in panels D to F (lowest
overall $R_{\rm LG}^2$) are roughly one order of magnitude smaller
compared to the other two data sets (panels A to F). This can be
explained by the fact that the corresponding initial reactant state is
characterized by $E_{\text{trans}}=0.5$ eV, which results in a low
reaction probability and renders a reactive collision a ``rare''
event. Consequently, the uncertainty arising from finite sample
statistics in the QCT simulations is largest for such data
sets. Moreover, 7 data sets with $E_{\text{trans}}=0.5$ eV had already
been excluded from the data set prior to training the NN because the
reaction probability obtained from QCT was negligible. This naturally
biases the NN training and predictions towards data sets with a larger
reaction probability.\\

\noindent
The product state distributions shown in Figure \ref{fig:fig5}
demonstrate the variety of shapes and features that are present. This
is a major difference compared to the product state distributions that
were considered for the DTD models. There, only $P(v')$ was subject to
significant variations, whereas $P(E_{\text{trans}}')$ and $P(j')$
showed less variability. Even for $P(v')$, three major classes of
distributions could be distinguished which is not the case for
STD. This variability explains the need for a denser grid and a more
expressive NN in the present work.\\

\begin{figure}
    \centering \includegraphics{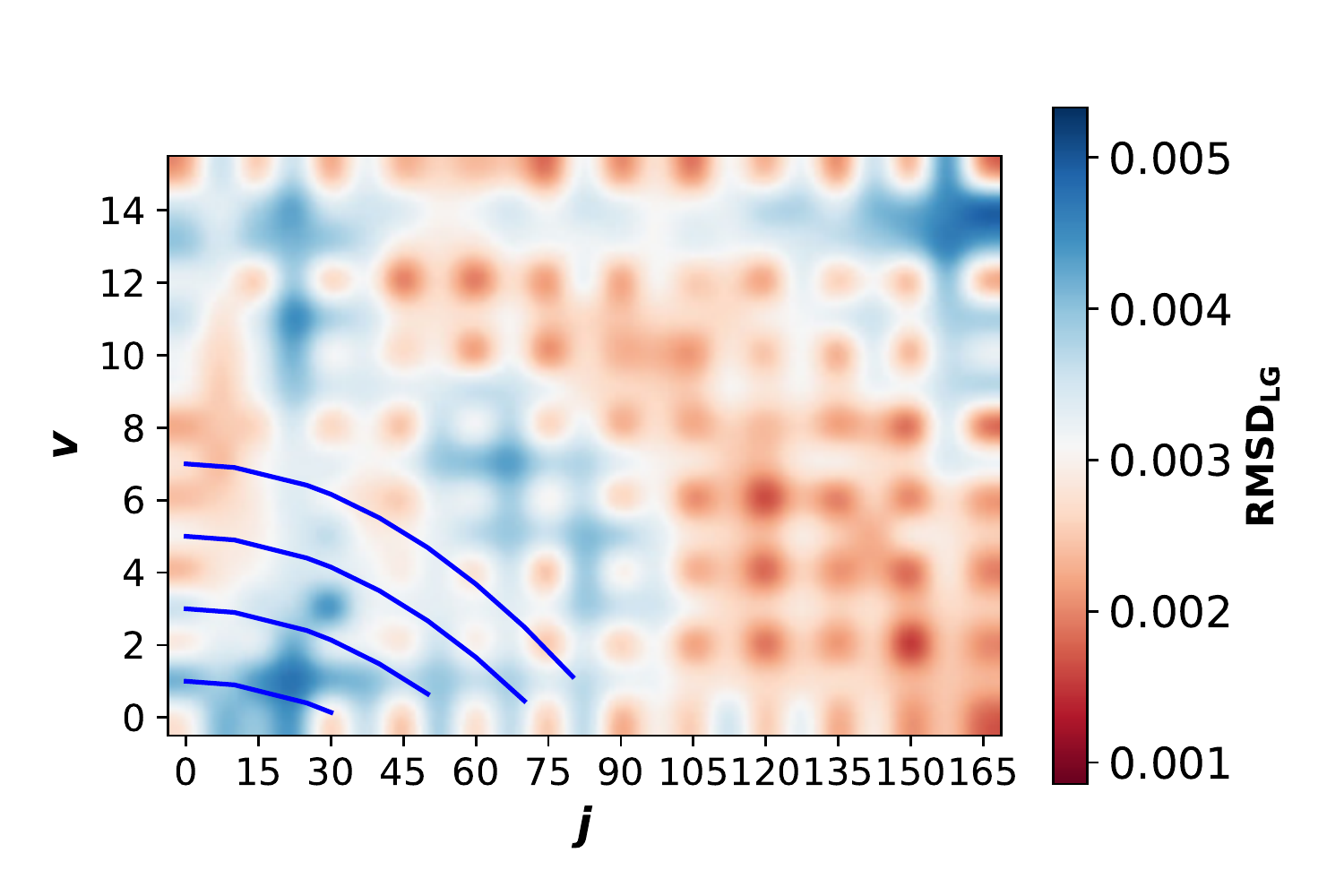}
    \caption{2D map for ${\rm RMSD_{\rm LG}}(v,j)$ between QCT and STD
      predictions for the product state distributions $P(E_{\rm
        int}^{'})$, $P(v')$, and $P(j')$ for given initial
      ($v,j$). The STD model was evaluated at fixed $E_{\rm
        trans}=4.0$ eV for the grid points $v=[0, 2, \cdots, 14]$ and
      $j=[0, 15, \cdots, 165]$ used for training, validation, and
      test. For the off-grid points, the $(v,j)-$combinations included
      $v=[1, 3, \cdots, 15]$ and $j=[7, 22, \cdots, 157]$. The solid
      blue lines indicate constant rovibrational energies of 0.2850
      eV, 0.6552 eV, 1.0142 eV, and 1.3622 eV. For the initial
      condition $(v=13, j=157)$ a comparatively high RMSD $(\sim
      0.005)$ is obtained (blue). For the corresponding $R^2_{\rm LG}$
      map, see Figure \ref{sifig:2dr2}.}
    \label{fig:fig6}
\end{figure}

\noindent
Next, the performance of the STD model on a larger grid including
parts of the training, test, validation set and additional initial
$(v,j)$ combinations is considered. For this, QCT simulations were
carried out for $v \in [0,15]$ with $\Delta v = 1$ and for $j =
[0,7,15,22,...,157,165]$. The entire grid considered included 368
points and 50000 QCT simulations for every $(v,j)$ combination were
run at $E_{\rm trans}=4.0$ eV. Figure \ref{fig:fig6} reports the ${\rm
  RMSD}_{\rm LG}$ between the product state distributions obtained
from QCT and those predicted by the STD model. The two-dimensional
surface ${\rm RMSD}_{\rm LG}(v,j)$ (for $R^2_{\rm LG}$ see
Figure~\ref{sifig:2dr2}) exhibits a visible checkerboard pattern that
reflects states $(v,j)$ used for training (on-grid) and off-grid
points which were not included in the training. Across the entire
$(v,j)$ state space the performance of STD is good. Despite the low
overall RMSD$_{\rm LG}$, there are regions (blue) that are associated
with larger differences between the reference QCT amplitudes and those
from the STD model. For low $(v,j)$ one reason for the somewhat larger
RMSD$_{\rm LG}$ is the low reaction probability whereas for high
$(v,j)$ neglecting ro-vibrational coupling may lead to increased
errors. A comparison of the final state distributions from QCT and the
STD model for $(E_{\rm trans}=4.0$ eV, $v=13, j=157)$ and $(E_{\rm
  trans}=4.0$ eV, $v=1, j=22)$ is reported in Figures
\ref{sifig:2dcomp} and \ref{sifig:2dcomplow}, respectively. A similar
deterioration of performance in the high temperature regime was, for
example, found from the surprisal model applied to the N$_2$+N
reaction\cite{schwartz:2018}.\\

\subsection{Performance for Initial Conditions from Reactant State Distributions}
Next, the ability of the STD model to predict product state
distributions given initial reactant state distributions is
assessed. For this, the STD model is tested for different types of
initial conditions by comparing reference product state distributions
from explicit QCT simulations with those predicted by the model. For a
given set of initial reactant state distributions initial conditions
($E_{\rm trans}, v, j$) are generated through Monte Carlo sampling. In
the limit of a sufficient number of samples, the average of the
product state distributions predicted by the STD model will converge
to the product state distributions associated with the given reactant
state distributions. Sampling 10000 initial conditions is sufficient
to converge the product state distributions obtained from STD (see
Figure \ref{sifig:convergence}A). This compares with $\sim 10^6$ that
are required for QCT simulations shown in Figure
\ref{sifig:convergence}B. The decrease in the number of samples
required for convergence is due to the more coarse-grained nature of
the STD model compared to QCT as the STD model lacks state-to-state
specificity for the products.\\

\noindent
Four distinct cases of thermal distributions are considered in the
following: $T_{\rm trans}=T_{\rm vib}=T_{\rm rot}$, $T_{\rm trans}
\neq T_{\rm vib} = T_{\rm rot}$, $T_{\rm trans} = T_{\rm vib} \neq
T_{\rm rot}$, and $T_{\rm trans} \neq T_{\rm vib} \neq T_{\rm
  rot}$. The performance measures of STD evaluated for the four cases
are summarized in Table~\ref{tab:4}. In all cases the STD model
provides an accurate prediction of product state distributions given
thermal reactant state distributions with ${\rm RMSD}_{\rm FG} \approx
0.003$ and $R^2_{\rm FG} \approx 0.996$. No significant differences in
STD model performance for the different cases is observed which
demonstrates that the STD model is generic in nature and applicable to
reactant state distributions of arbitrary shape with significant
weight over the range of initial reactant states considered in
training. The decreased level of performance for predicting $P(E_{\rm
  int}^{'})$ distributions compared to $P(v')$ or $P(j')$ is again
attributed to stronger variation in shapes and peaks near the NO
dissociation as was already found for final state distributions from
individual initial reactant states, see Table \ref{tab:1}. Moreover,
the cutoff at $E_{\rm trans}=8.0$ eV in the training data of the STD
model becomes relevant for $P(E_{\rm int}^{'})$ distributions at high
temperatures and may lead to a decrease in performance.\\

\begin{table}[ht]
\footnotesize \centering
\begin{tabular}{l|cccc|cccc}
\hline
\hline

& \multicolumn{4}{c|}{RMSD$_{\rm FG}$} & \multicolumn{4}{c}{$R_{\rm FG}^{2}$} \\
\hline
& Overall & $P(E_{\rm int}^{'})$ & $P(v')$ & $P(j')$  & Overall & $P(E_{\rm int}^{'})$ & $P(v')$ & $P(j')$  \\
\hline
$T_{\rm trans}=T_{\rm vib}=T_{\rm rot}$& 0.0029 & 0.0079 & 0.0007 & 0.0001  & 0.9965 & 0.9912 & 0.9993 & 0.9990  \\
$T_{\rm trans} \neq T_{\rm vib} = T_{\rm rot}$ & 0.0028  & 0.0078 & 0.0007 & 0.0001 & 0.9961 & 0.9904 & 0.9990 & 0.9990  \\
$T_{\rm trans} = T_{\rm vib} \neq T_{\rm rot}$ & 0.0030 & 0.0083 & 0.0007 & 0.0001 & 0.9961 & 0.9900 & 0.9992 & 0.9990  \\
$T_{\rm trans} \neq T_{\rm vib} \neq T_{\rm rot}$ & 0.0030 &  0.0081 & 0.0007 & 0.0001 & 0.9953 & 0.9885 & 0.9985 & 0.9989\\
\hline
\hline
\end{tabular}
\caption{Performance comparison of the STD model in terms of {${\rm
      RMSD}_{\rm FG}$} and $R^2_{\rm FG}$ for the 4 different
  temperature sets: $T_{\rm trans}=T_{\rm vib}=T_{\rm rot}$ (61 sets),
  $T_{\rm trans} \neq T_{\rm vib} = T_{\rm rot}$ (3637 sets), $T_{\rm
    trans} = T_{\rm vib} \neq T_{\rm rot}$ (60 sets), and $T_{\rm
    trans} \neq T_{\rm vib} \neq T_{\rm rot}$ (840 sets).}
\label{tab:4}
\end{table}

\begin{figure}[h!]
\centering
\includegraphics[scale=0.6]{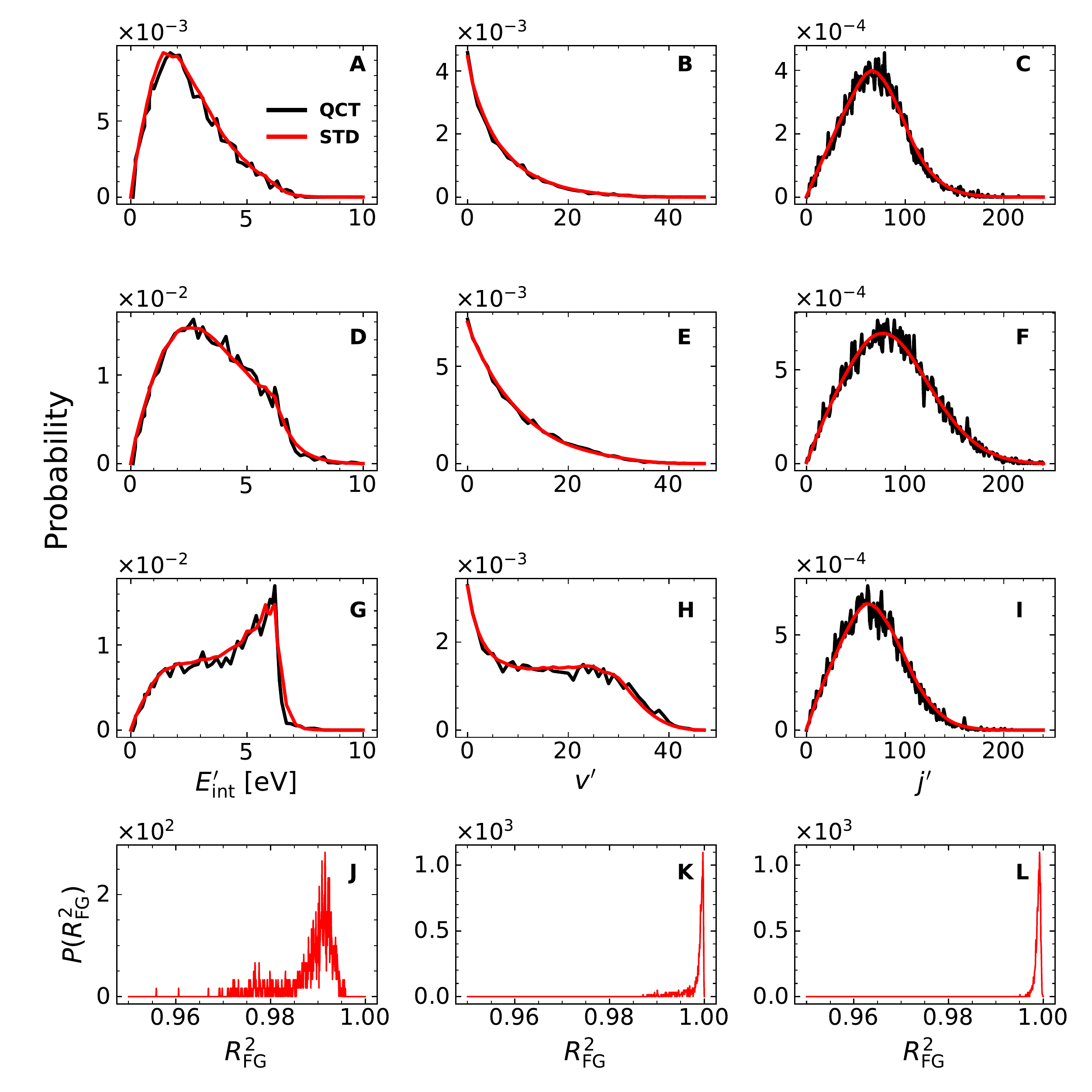}
\caption{Product state distributions for $T_{\rm trans} \neq T_{\rm
    rot} \neq T_{\rm vib}$ obtained from STD and compared with
  explicit QCT simulations. Panels A-C: best performing prediction
  (largest $R^{2}_{\rm FG}$) for $T_{\rm trans}= 10000 $ K, $T_{\rm
    vib}=6000$ K , $T_{\rm rot}=5000$ K; panels D-F: prediction
  closest to the mean $R^{2}_{\rm FG}$ for $T_{\rm trans}= 5000 $ K,
  $T_{\rm vib}=8000$ K, $T_{\rm rot}=13000$ K; panels G-I: worst
  performing model (smallest $R^{2}_{\rm FG}$) value for $T_{\rm
    trans}= 5000 $ K, $T_{\rm vib}=18000$ K, $T_{\rm rot}=8000$
  K. Panels J-L: normalized distributions $P(R_{\rm FG}^{2})$ for the
  complete set of 840 temperatures ($T_{\rm trans} \neq T_{\rm rot}
  \neq T_{\rm vib}$), respectively, (J) $P(E_{\rm int}^{'})$, (K)
  $P(v')$ and (L) $P(j')$.}
\label{fig:fig7}
\end{figure}

\noindent
For the most general case $T_{\rm trans} \neq T_{\rm rot} \neq T_{\rm
  vib}$, 840 temperature combinations were generated. As
Figure~\ref{fig:fig7} demonstrates the STD model reliably captures
overall shapes and features such as the position of maxima even for
the worst performing data set (panels G to I). This is remarkable as
the shapes of $P(E_{\rm int}^{'})$ and $P(v')$ can vary
appreciably. The distribution of $R^2$ values for all 840 data sets
also demonstrates high prediction accuracy, in particular for $P(v')$
and $P(j')$. The specific case $T_{\rm trans} \neq T_{\rm vib} =
T_{\rm rot}$ is considered in Figure \ref{sifig:STD-DTD-set1}. Panels
\ref{sifig:STD-DTD-set1}A to C are for the best performing STD model
compared with QCT data whereas panels D to F are representative for
the average $R_{\rm FG}^2$. Both examples demonstrate that shapes and
location of maxima are reliably captured by predictions based on the
STD model. Even for the worst performing STD model (panels G to I) the
important features of the distributions are still captured
reliably. Finally, Figures \ref{sifig:STD-DTD-set1}J to L report the
distribution $P(R_{\rm FG}^2)$ for all 3637 models evaluated for
$T_{\rm trans} \neq T_{\rm vib} = T_{\rm rot}$. For all distributions
$R_{\rm FG}^2 > 0.95$ with $P(j')$ performing best.\\

\noindent
A direct comparison of the STD and DTD models is reported in
Table~\ref{tab:2}. The two models perform on par for all measures and
all degrees of freedom except for $E_{\rm int}'$. This is despite the
fact that the DTD model was explicitly trained on these thermal
distributions (4658 data sets in total) and further underlines the
predictive power of the STD model. Also, it should be noted that for
the DTD model $E_{\rm trans}'$ instead of $E_{\rm int}'$ was used for
training. Given the excellent performance of both models, the
differences appear to be negligible. As such, STD represents a highly
accurate approach to obtain product state distributions given initial
state specific reaction states. The decreased level of performance for
predicting $P(E_{\rm int}^{'})$ distributions compared to $P(v')$ or
$P(j')$ (see RMSD$_{\rm FG}$ and $R_{\rm FG}^2$) has several
origins. First, $P(E_{\rm int}^{'})$ distributions vary strongly in
shape and are typically peaked (see Figure \ref{fig:pofe}) which is
challenging to capture using a G-based approach. Secondly, for highly
excited $(v',j')$ states rovibrational coupling in the diatomic
product molecule becomes more important. Explicitly accounting for
this coupling during the data preparation may further improve the
predictive power of STD. In contrast to $P(E_{\rm int}^{'})$,
rotational distributions $P(j')$ vary least and can thus be predicted
with the highest accuracy. Moreover, the cutoff at $E_{\rm trans}=8.0$
eV in the training data of the STD becomes relevant for $P(E_{\rm
  int}^{'})$ distributions at high temperatures and may lead to a
decrease in performance. The small decrease in accuracy when comparing
${\rm RMSD}_{\rm LG}$ and $R^{2}_{\rm LG}$ to ${\rm RMSD}_{\rm FG}$
and $R^{2}_{\rm FG}$, see Table \ref{tab:2}, arises because of the
linear interpolation to obtain predicted amplitudes between the
designated grid points.\\

\begin{table}[ht]
\footnotesize \centering
\begin{tabular}{l|cc|cc}
\hline
\hline

& \multicolumn{2}{c|}{RMSD$_{\rm FG}$} & \multicolumn{2}{c}{${R_{\rm FG}^2}$}   \\
\hline
&STD|QCT &  DTD|QCT & STD|QCT & DTD|QCT \\
\hline
overall& 0.0030 & 0.0017 & 0.9953 & 0.9988 \\
$E_{\alpha}^{'}$& 0.0081 & 0.0042 & 0.9885 & 0.9985 \\
$v'$& 0.0007 & 0.0008 & 0.9985 & 0.9988 \\
$j'$& 0.0001 & 0.0001 & 0.9989 & 0.9991  \\
\hline
\hline
\end{tabular}
\caption{Performance of STD and DTD models for $T_{\rm rot} \neq
  T_{\rm vib}$ (960 data sets) compared with QCT results for initial
  conditions from initial thermal distributions. For the STD model
  $\alpha = {\rm int}$ and for DTD $\alpha = {\rm trans}$.
  Performance measures (averaged over the all data sets) ${\rm
    RMSD}_{\rm FG}$ and ${R}_{\rm FG}^2$ are computed by comparing QCT
  data with the STD or DTD model predictions over the grid for which
  explicit QCT data is available. For $T_{\rm trans}= 5000, 10000,
  15000, 20000$ K a set of 960 temperatures is evaluated with $T_{\rm
    rot} \neq T_{\rm vib}$ ranging from 5000 to 20000 K with $\Delta T
  = 1000$ K.}
\label{tab:2}
\end{table}

\noindent
It is also of interest to compare the performance of STD in predicting
QCT data with the fidelity of the QCT data itself. As training of the
NN is based on final state distributions from $8 \times 10^4$
trajectories for each initial condition it is likely that the training
set does not contain fully converged reference information. To this
end, a much larger number $(N_{\rm C} = 5 \times 10^6)$ of QCT
simulations was carried out for a few initial conditions to determine
the ``ground truth'' and were compared with final state distributions
from only $N_{\rm U} = 5 \times 10^4$ samples. The correlation for the
bin-occupation between the ``ground truth'', i.e. ``converged''
distributions from $N_{\rm C}$ samples, and the unconverged
distributions using $N_{\rm U}$ samples is $R^2 \sim 0.99$ or better
for all four initial conditions considered and all three degrees of
freedom, see Figure~\ref{sifig:qctconv}. Hence, the quality of the QCT
reference data used for training the NN is comparable to the
performance of the NN itself. The relative error between distributions
from ``ground truth'' and the unconverged samples is around 0.2, see
Figure \ref{sifig:qctvalid}. Thus, the $\chi^2$ (rescaled mean squared
error) between STD and reference QCT simulations accounting for the
fact that the QCT input to train the NN is not fully converged is
about 5 times larger than the RMSD, which yields $\chi^2 \sim
0.005$. One possible way of looking at this is to consider the amount
of information (or signal) compared to the amount of noise. This
``signal-to-noise ratio'' should increase $\propto \sqrt{N}$ where $N$
is the number of samples, assuming that the noise is stochastic and
arising from insufficient sampling. As seen in Figure
\ref{sifig:qctvalid}, when the number of samples in a channel is above
$\sim 10$, the noise/signal is 0.1.\\

\noindent
The relevance of ``rare events'' is a major difference when
considering product state distributions from individual initial
reactant states compared to initial conditions from reactant state
distributions. When applied to individual initial conditions it was
found that the STD model performance decreases for $E_{\rm trans} \leq
1.0$ eV, i.e. for initial conditions with low reaction
probability. The corresponding product state distributions are noisy
and show large variations due to finite sample statistics from
QCT. This may be improved in future work through importance sampling
of the impact parameter. While rare events are crucial for an accurate
description of certain physical phenomena, such as plasma
formation,\cite{piel:2017} they do {\it not} constitute a significant
contribution to product state distributions. As such, for observables
that involve integration of a product state distribution, such as
reaction rates, the decrease of performance of the STD model with
regards to rare events is also negligible.\\

\begin{figure}[h!]
\centering \includegraphics[scale=0.4]{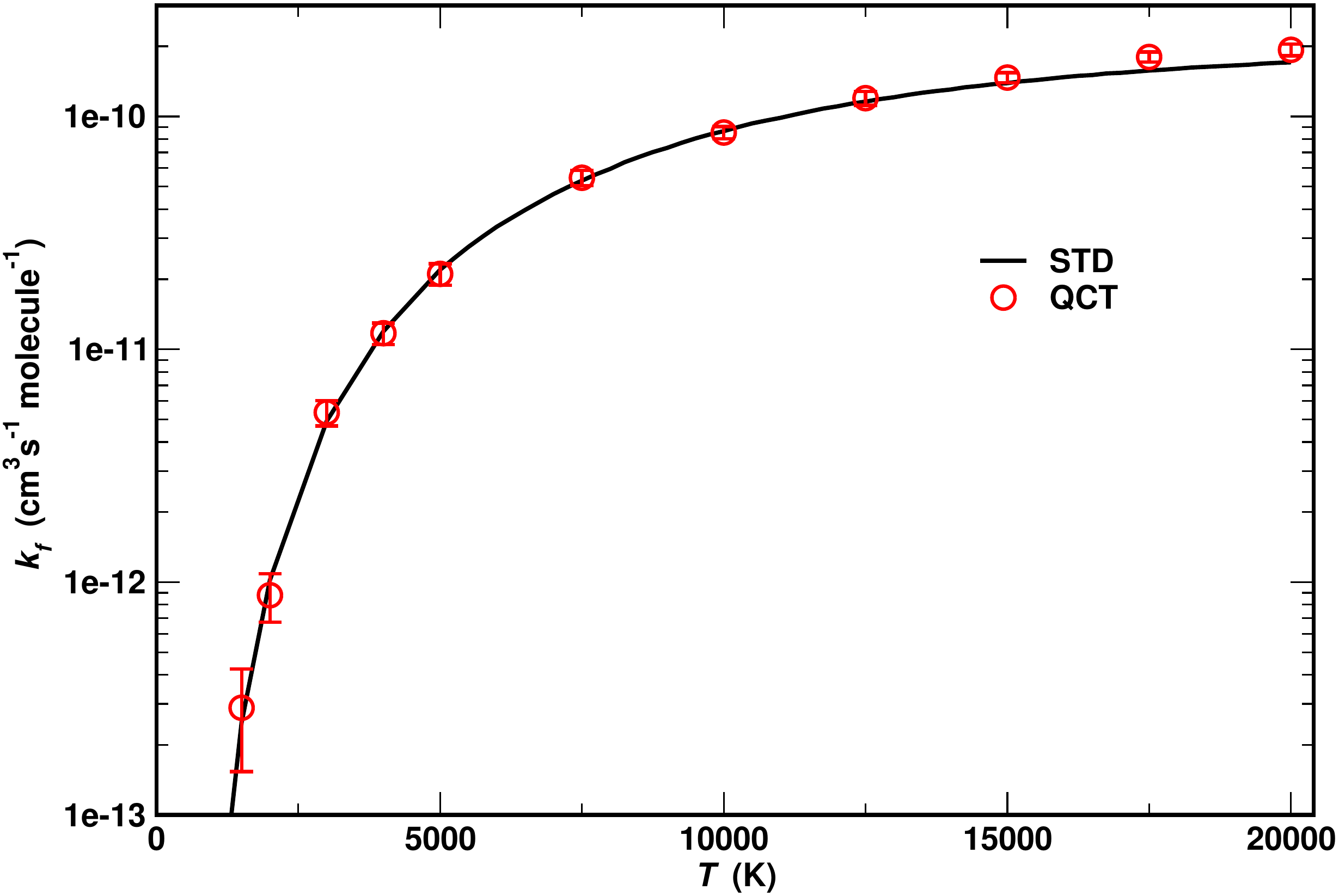}
\caption{The thermal forward rate $k_f$ calculated from QCT (open red
  circle) and STD model (solid black line) for the $^4$A$'$ state of
  the N($^4$S)+O$_{2}$(X$^3 \Sigma_{\rm g}^{-}$) $\rightarrow$
  NO(X$^2\Pi$) +O($^3$P) reaction between 1500 and 20000 K. The
  present rates agree quantitatively with those directly obtained from
  QCT simulations.\cite{MM.no2:2020} It is interesting to note that
  significant differences between the rates from QCT simulations and
  those from the STD model arise only for the highest temperatures for
  which pronounced $v-j$ coupling is expected.}
\label{fig:figrate}
\end{figure}

\noindent
From the STD-predicted product state distributions, $T-$dependent
reaction rates can be obtained and compared with rates from explicit
QCT simulations. In general, such a rate is determined from
\begin{equation}
    k(T)=g(T) \sqrt{\frac{8k_{\rm B}T}{\pi \mu}}\pi b^{2}_{\rm max} P_{\rm r},
\end{equation}
where $P_{\rm r}$ is the probability for a reaction to occur. For QCT
simulations $P_r = \frac{N_{\rm r}}{N_{\rm tot}}$ where $N_{\rm r}$ is
the number of reactive trajectories and $N_{\rm tot}$ is the total
number of trajectories run. For the STD model, $P_{\rm r} =
\int_{E=0}^{E_{\rm max}} P(E)dE$ where $E = E_{\rm int}'$. For the
forward N($^4$S)+O$_{2}$(X$^3 \Sigma_{\rm g}^{-}$) $\rightarrow$
NO(X$^2\Pi$) +O($^3$P) reaction on the $^4$A$'$ electronic state the
degeneracy factor $g(T) = 1/3$ and $\mu$ is the reduced mass of the
reactants.\cite{MM.no2:2020} The two approaches are compared in Figure
\ref{fig:figrate} and favourable agreement is found over a wide
temperature range. Hence, the STD model can also be used to determine
macroscopic quantities such as realistic reaction rates which is
essential. The decrease in prediction accuracy at the highest
temperatures may be attributed to the cutoff at $E_{\rm trans}=8.0$ eV
in the training data of the STD. Cross sections $\sigma = \pi b_{\rm
  max}^2 P_{\rm r}$ were also determined for the test set $(N_{\rm
  test} = 77)$. Typical values for $\sigma$ from the QCT simulations
are $\sigma \sim 9\times 10^{-15}$ cm$^2$ which compares with those
from the STD model of $\sigma \sim 8.5\times10^{-15}$ cm$^2$.\\

\noindent
Finally, it is also of interest to compare the computational cost for
evaluating the STD and DTD models. Here, a single evaluation refers to
the prediction of the product state distributions at 201, 48, and 241
evenly spaced points for $E_{\rm int}'$ (for STD) or $E_{\rm trans}'$
(for DTD) between 0 and 20 eV, $v'=0-47$, and $j'=0-240$,
respectively, for a given reactant state distribution. The evaluation
time for processing 50 reactant state distributions randomly selected
from the total set of 4658 distributions is ($98.01 \pm 5.95 $) s
using the STD model and ($1.03 \pm 0.01 $) s using the DTD model on a
3.6 GHz Intel Core i7-9700K CPU. The difference of two orders of
magnitude is explained as follows. For the STD model the NN is
considerably larger and STD requires 10000 NN-evaluations for a given
reactant state distribution. Contrary to that, for DTD only one
evaluation is required. On the other hand, for STD linear
interpolation is used to obtain an amplitude whereas DTD needs to
evaluate a computationally costly kernel-based interpolation.\\

\section{Discussion and Conclusion}
The present work introduces a machine-learned state-to-distribution
model for predicting final state distributions from specific initial
states of the reactants. The STD model achieves a good performance,
see Tables \ref{tab:1} and \ref{tab:2}, and accurately predicts
product state distributions as compared with reference QCT
simulations. The model also allows to determine observables such as
thermal reaction rates, see Figure \ref{fig:figrate}.\\

\begin{figure}[bth!]
\begin{center}
\includegraphics[width=0.99\textwidth]{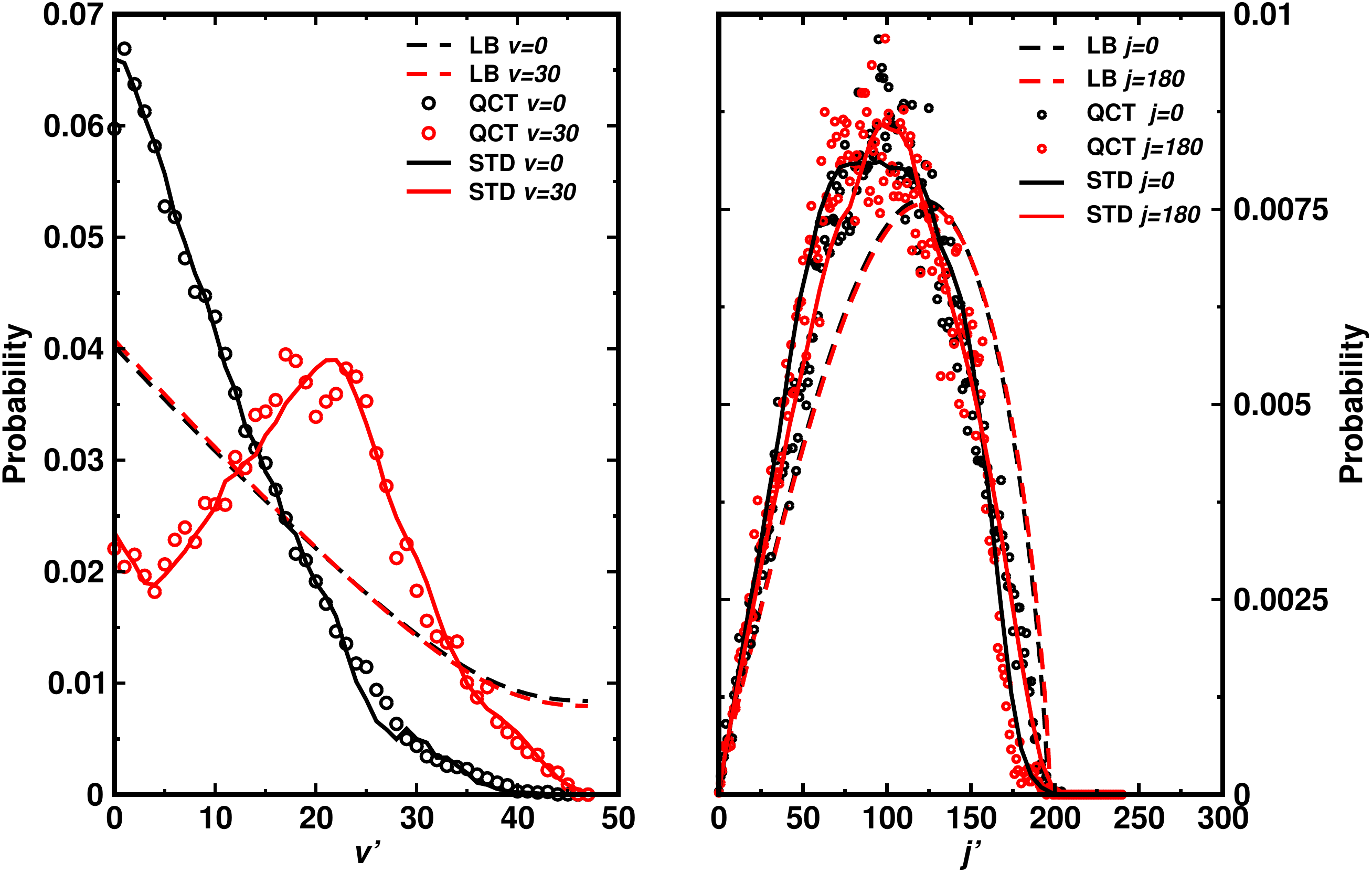}
\caption{Comparison of the final state distributions from the STD
  model and those from the Larsen-Borgnakke model which is often used
  in DSMC simulations. The initial conditions used are: ($E_{\rm
    trans}=7.0$ eV $v=0,j=0);(E_{\rm trans}=2.5$ eV,$v=30,j=0);(E_{\rm
    trans}=2.0$ eV$,v=0,j=180$), all at the same $\varepsilon_{\rm
    coll}$ or $E_{\rm tot}$.}
\label{fig:fig9}
\end{center}
\end{figure}

\noindent
One specific motivation to develop such an STD model is for generating
meaningful input for direct simulation Monte
Carlo~\cite{bird1976molecular} (DSMC) simulations. DSMC is a
computational technique to simulate nonequilibrium high-speed flows
and is primarily applied to dilute gas flows. The method is
particle-based, where each particle typically represents a collection
of real gas molecules, and transports mass, momentum, and
energy. Models are required to perform collisions between particles by
which they exchange momentum and energy with one another. For
instance, given the internal energy states and relative translational
energy of reactants in a colliding pair of particles, the total
collision energy (TCE) model proposed by Bird is a widely used
quantity to estimate the reaction
probability~\cite{bird1976molecular}. Once a colliding pair is
selected for a collision, a key model output is the post-collision
energy distribution from which product states are subsequently
sampled. The state-of-the-art model for such a purpose is a
phenomenological model proposed by
Larsen-Borgnakke~\cite{borgnakke1975statistical} (LB). The explicit
form of the LB model for sampling the rotational and vibrational
energy after a reaction is
\begin{equation}
    f_{\rm LB} =
    \cfrac{\left[1-\cfrac{\varepsilon_{i}'}{\varepsilon_{coll}}
        \right]^{\zeta_{\rm tr}/2-1}}{\sum_i
      \left[1-\cfrac{\varepsilon_{i}'}{\varepsilon_{coll}}
        \right]^{\zeta_{\rm tr}/2-1}},
\label{eq:lb}
\end{equation}
where $\varepsilon_i'$ corresponds to $\varepsilon_v'$ or
$\varepsilon_j'$ for post-reaction vibrational and rotational energy
respectively and $\varepsilon_{\rm coll}=\varepsilon_{\rm
  v}'+\varepsilon_{\rm j}'+\varepsilon_{\rm t}'$. That is, the
collision energy is the sum of the internal energy and translational
energy post-collision (or pre-collision, due to conservation of total
energy in the system the two coincide). In Eq. \ref{eq:lb},
$\zeta_{\rm tr} = 5-2\omega$ is related to the translational degrees
of freedom and the collision cross section parameter $\omega$ which is
obtained by fitting collision cross sections such that the viscosity
$\mu \propto T^\omega$.  of a gas is recovered. In essence, the LB
model is not based on state-specific probabilities of product states
in reactions. Figure \ref{fig:fig9} reports the LB model results
together with predictions from the STD model and reference QCT
simulations. The significant discrepancies are not surprising as the
LB model samples post-collision states from a local equilibrium
distribution. This is the advantage of the STD model which is based on
state-specific reference calculations from QCT simulations.\\

\noindent
One additional refinement of the present method concerns preparation
of the data set for training the NN. Including rotation/vibration
coupling is likely to improve the overall model specifically for high
$(v,j)$ states. Furthermore, generating initial conditions from
stratified sampling of the impact parameter may more broadly cover
low-energy initial translational energies to further extend the range
of applicability of the trained NN.\\

\noindent
In conclusion, an initial state-resolved model to predict final state
distributions for chemical reactions of type A+BC$\rightarrow$AB+C
based on machine learning is formulated and tested. The prediction
quality of the model compared with explicit QCT simulations is
characterized by ${\rm RMSD} \sim 0.003$ and $R^2 \sim 0.99$. Final
state distributions from STD can be sampled again using Monte Carlo
simulations for generating input for more coarse grained simulations,
such as DSMC. Furthermore, the STD model complements the DTD model
when predicting product from reactant state distributions. At the cost
of an increased evaluation time the STD model allows for accurate
predictions given arbitrary nonequilibrium reactant state
distributions. This is a regime for which DTD models trained on a
given set of (equilibrium) reactant state distributions may
underperform. In conjunction, these two models can enable the
efficient and accurate simulation of molecular systems over time
undergoing multiple reactive collisions.\\

\section*{Data and Code Availability}
Exemplary data sets and code for evaluating STD models is available at
\url{https://github.com/MMunibas/STD}.

\section*{Supporting Information}
The Supporting Information contains a detailed description of the
local averaging procedure together with
Figures~\ref{sifig:fig4}-\ref{sifig:qctvalid}.

\section{Acknowledgment}
This work was supported by AFOSR, the Swiss National Science
Foundation through grants 200021-117810, 200020-188724, the NCCR MUST,
and the University of Basel.

\bibliography{refs}

\clearpage

\renewcommand{\thetable}{S\arabic{table}}
\renewcommand{\thefigure}{S\arabic{figure}}
\renewcommand{\thesection}{S\arabic{section}}
\renewcommand{\d}{\text{d}}
\setcounter{figure}{0}  
\setcounter{section}{0}  

\noindent
{\bf Supporting Information: Machine Learning Product State
  Distributions from Initial Reactant States for a Reactive
  Atom-Diatom Collision System}\\

\section{Detection of ``sharp'' peaks}
\label{si_sec_peak_detection}
When constructing the training, validation, and test data from the raw
QCT data no local averaging was performed around "sharp'' peaks and
averaging over fewer points was done at nearby points. This was done
to conserve the sharp peaks, as they would otherwise be washed out. A
maximum of a given distribution was classified as ``sharp'' based on
the following criteria:

For $P(E_{\rm int}')$ if there were 2, 1, or 0 points to either side
of the maximum, it was classified as sharp. Otherwise two linear fits
were done to the 3 points (or 4 points, if available) to the left and
right of a maximum, respectively, including the maximum itself. If the
magnitude of both slopes exceeded a critical value $|a_{\rm crit}| =
0.001$, the maximum was classified as ``sharp''. Subsequently,
averaging over neighbouring data points was performed: The maximum was
not averaged, the nearest and next-nearest neighbours of the maxima
were averaged with $n_{\rm max} = 2$, and all other points were
averaged with $n_{\rm max} = 3$.\\

\begin{figure}[h!]
\begin{center}
\includegraphics[width=0.99\textwidth]{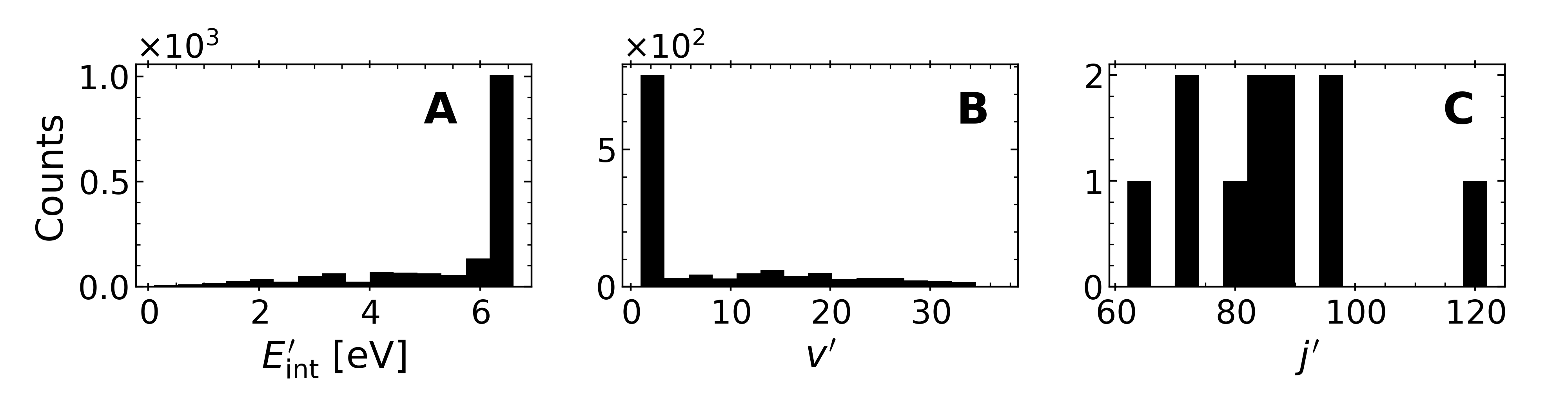}
\caption{Histograms showing the distribution of ``sharp'' peaks for
  (A) $P(E_{\rm int}^{'})$, (B) $P(v')$ and (C) $P(j')$ obtained by
  QCT simulations, considering all 2184 initial reactant states
  considered in the present work.}
\label{sifig:fig4}
\end{center}
\end{figure}

For $P(v')$ if there were 2, 1, or 0 points to either side of a
maximum, it was classified as sharp. Otherwise two linear fits were
performed for the 3 points to the left and right of the maximum
including the maximum itself. If the magnitude of both slopes exceeded
a critical value $|a_{\rm crit}| = 0.000143$, the maximum was
classified as ``sharp''. Subsequently, averaging over neighbouring
data points was performed: The maximum was not averaged, and all other
points were averaged with $n_{\rm max} = 1$. The same procedure
applied to $P(j')$ with $n_{\rm max} = 7$ and $|a_{\rm crit}| =
0.000005$. Figure~\ref{sifig:fig4} shows the distribution of ``sharp''
peaks for all 2184 explicit initial reactant states considered in this
work.
\section{Additional figures}

\begin{figure}[h!]
    \centering
    \includegraphics{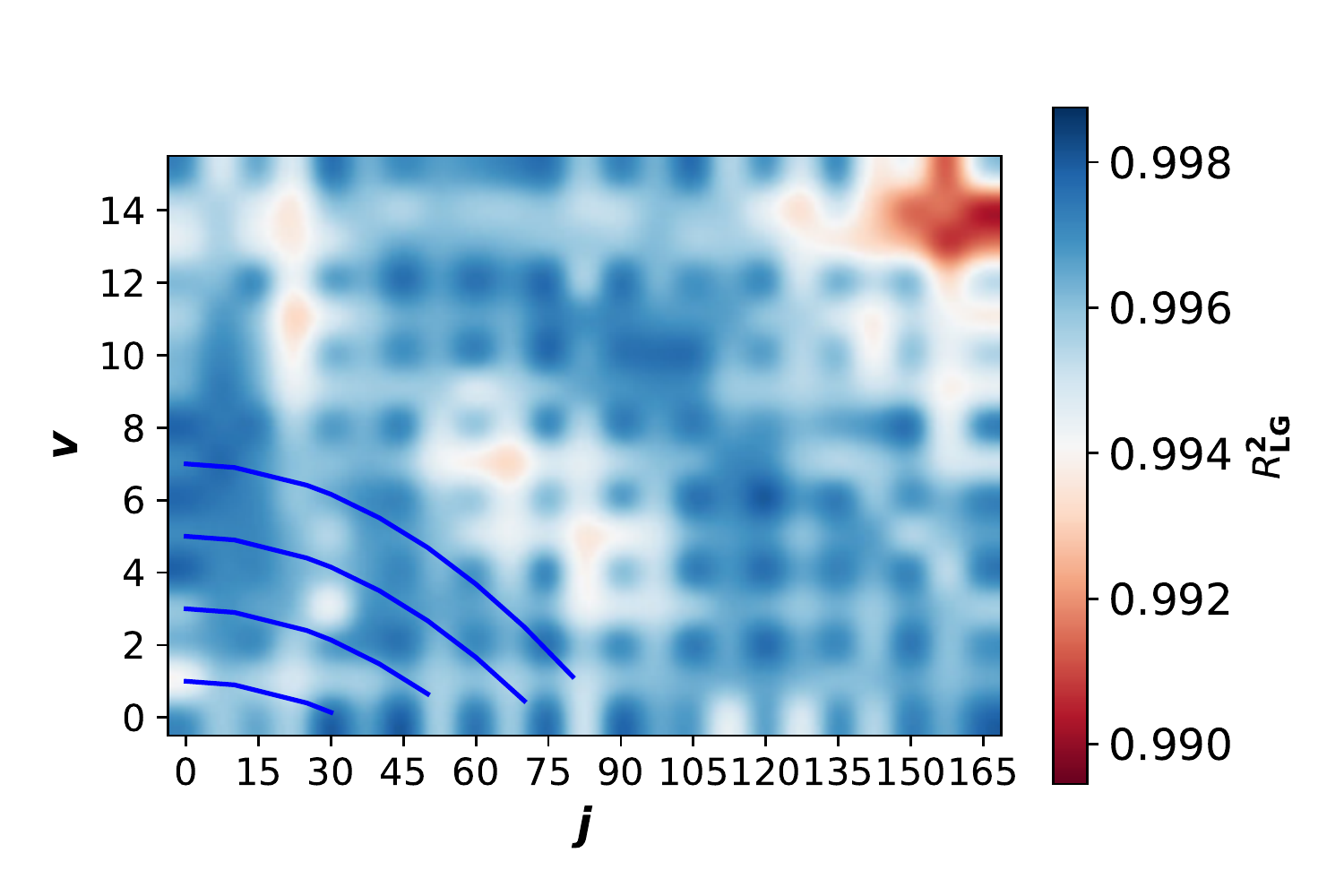}
    \caption{2D map of the $R^{2}_{\rm LG}$ values between QCT and STD
      predictions for the product state distributions $P(E_{\rm
        int}^{'})$ , $P(v')$, and $P(j')$ for given initial
      ($v,j$). The STD model was evaluated at fixed $E_{\rm
        trans}=4.0$ eV for the grid points $v=[0, 2, 4, 6, 8, 10, 12,
        14]$ and $j=[0, 15, 30, 45, 60, 75, 90, 105, 120,
        135,150,165]$ used for training, validation, and testing. For
      the offgrid points, the $(v,j)-$combinations included $v=[1, 3,
        5, 7, 9, 11,13,15]$ and $j=[7, 22, 37, 52, 67, 82, 97, 112,
        127, 142,157]$. The red lines are for constant rovibrational
      energies of 0.2850 eV, 0.6552 eV, 1.0142 eV, and 1.3622 eV. For
      the initial condition $(v=13, j=160)$ a low value of $R^2 =
      0.94$ is obtained (red); the direct comparison between the final
      state distributions from QCT and those predicted from STD is
      given in Figure \ref{sifig:2dcomp}.}
    \label{sifig:2dr2}
\end{figure}

\begin{figure}[h!]
  \centering \includegraphics[width=\textwidth]{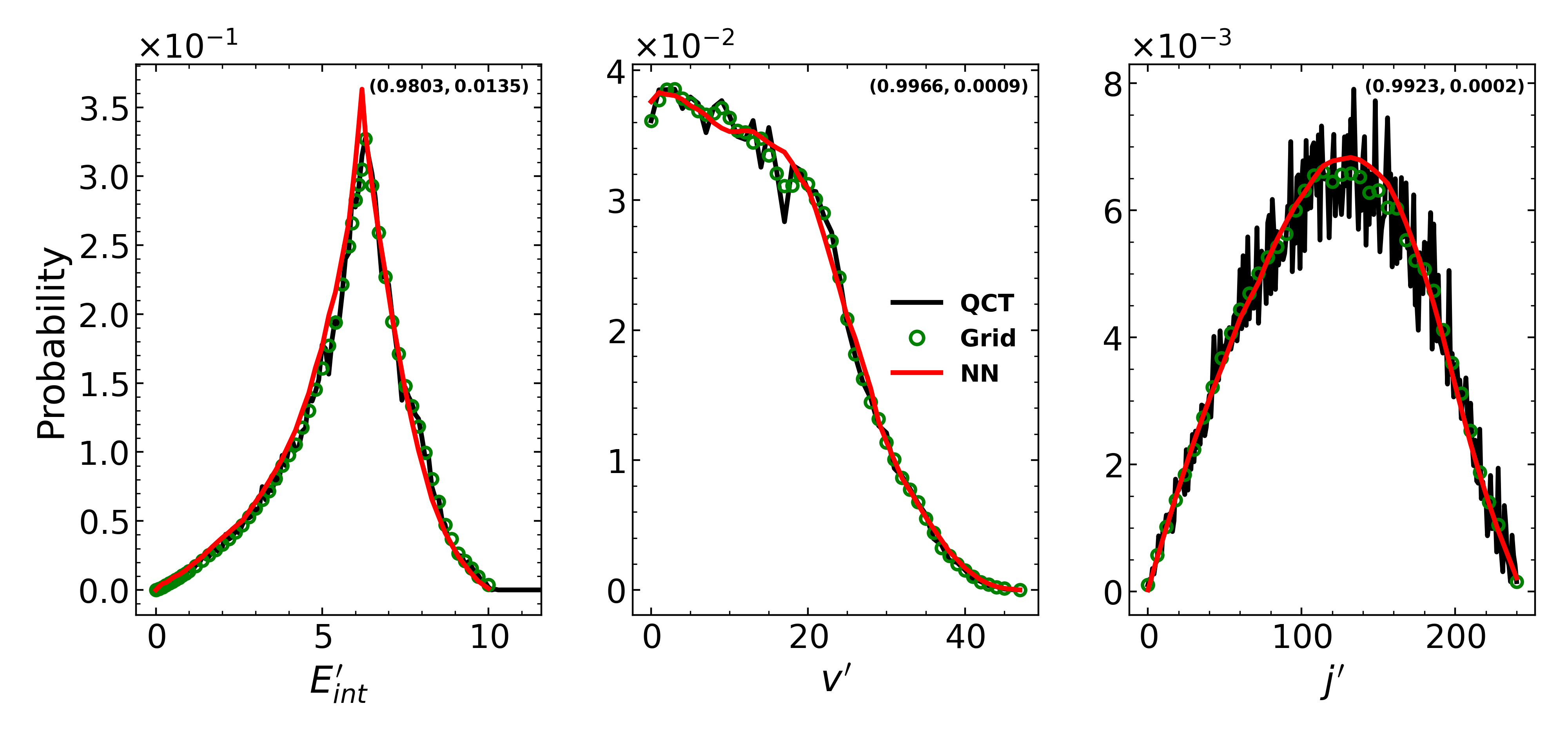}
    \caption{Direct comparison between final state distributions from
      QCT simulations (black solid line) and those predicted from the
      STD model (red solid line) for ($E_{\rm trans}=4.0$ eV, $v=13$,
      $j=157$). The $R^{2}_{\rm LG}$ and ${\rm RMSD}_{\rm LG}$ for
      $P(E_{\rm int}')$, $P({v'})$ and $P({j'})$ are shown in
      parenthesis.  The overall $R^{2}_{\rm LG}$ and ${\rm RMSD}_{\rm
        LG}$ are 0.9897 and 0.0049, respectively.}
    \label{sifig:2dcomp}
\end{figure}

\begin{figure}[h!]
  \centering \includegraphics[width=\textwidth]{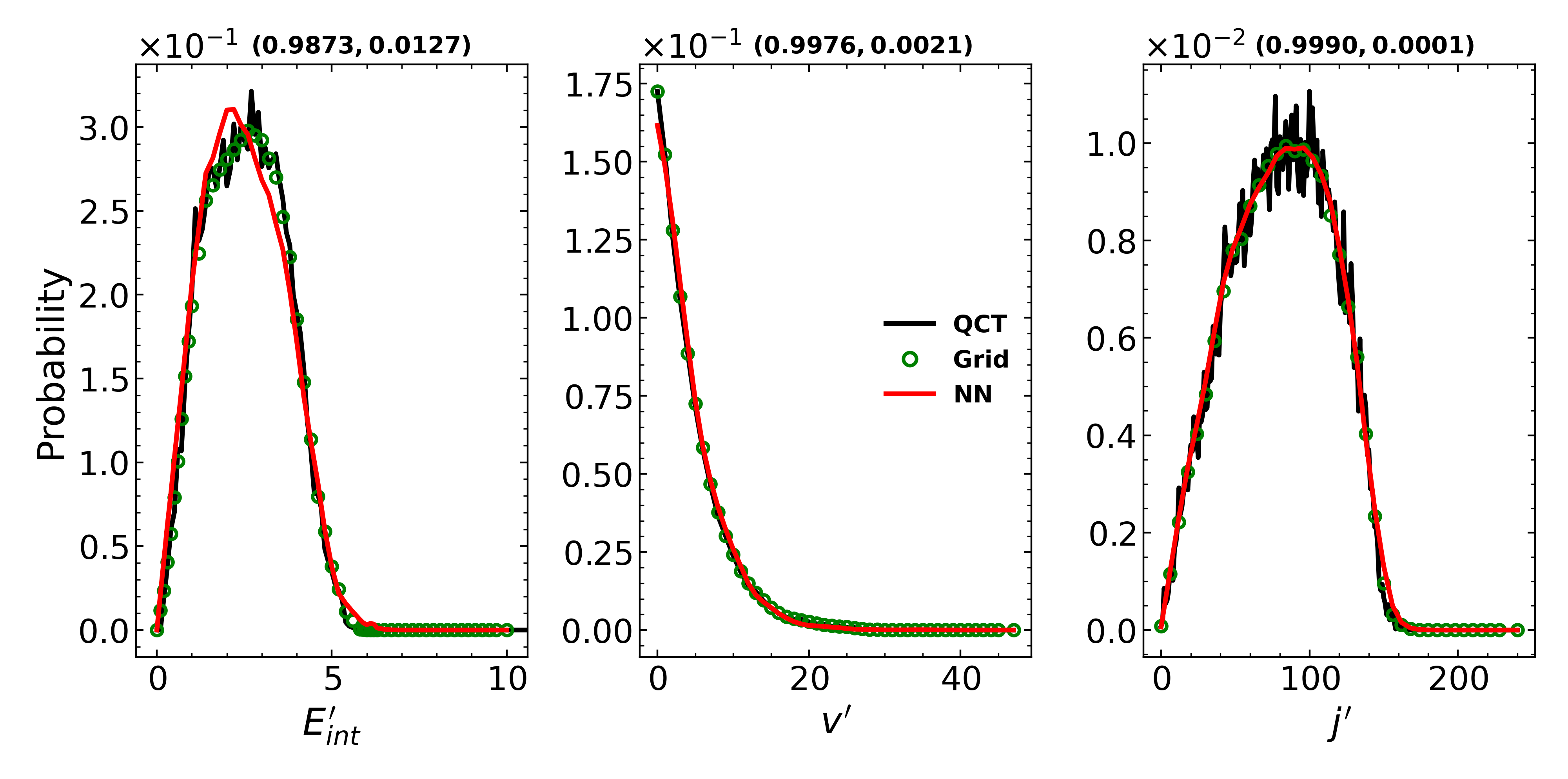}
    \caption{Direct comparison between final state distributions from
      QCT simulations (black solid line) and those predicted from the
      STD model (red solid line) for initial condition ($E_{\rm
        trans}=4.0$ eV, $v=1$, $j=22$). The $R^{2}_{\rm LG}$ and ${\rm
        RMSD}_{\rm LG}$ for $P(E_{\rm int}')$, $P({v'})$ and $P({j'})$
      are shown in parenthesis.  The overall $R^{2}_{\rm LG}$ and
      ${\rm RMSD}_{\rm LG}$ are 0.9946 and 0.0050, respectively.}
    \label{sifig:2dcomplow}
\end{figure}

\begin{figure}[h!]
    \centering \includegraphics[scale=0.3]{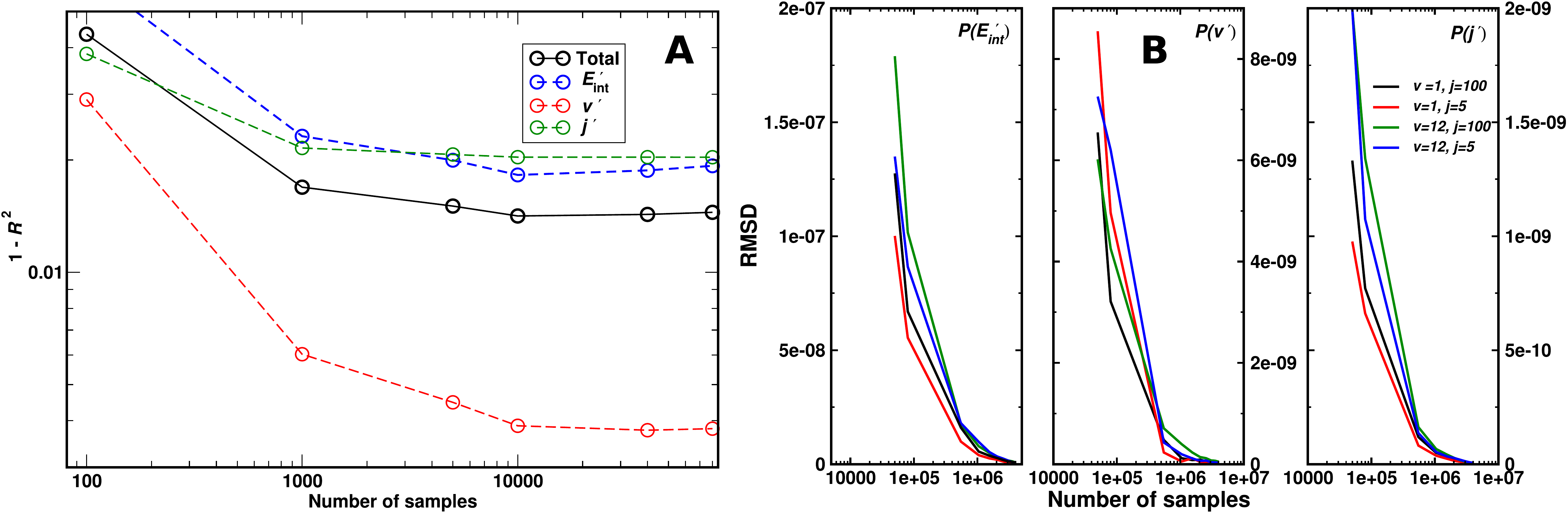}
    \caption{Panel A: Error of the STD model in predicting product
      state distributions as a function of the number of samples drawn
      from the reactant state distributions. The error is reported as
      $1-R^2_{\rm FG}$, where $R^2_{\rm FG}$ is obtained by averaging
      over the entire set of temperatures with $T_{\rm trans}\neq
      T_{\rm vib} \neq T_{\rm rot}$. The error saturates at $\approx
      10000$ samples. Panel B: RMSD of QCT data as a function of
      different trajectories sample size with ($5 \times 10^{6}$)
      trajectories as reference.  Four different initial conditions
      ($v=1, j=100; v=1, j=5; v=12, j=100; v=12, j=5$) at $E_{\rm
        trans}= 4.0$ eV are evaluated for $P(E_{\rm int}')$, $P(v')$,
      and $P(j')$.}
    \label{sifig:convergence}
\end{figure}

\begin{figure}[h!]
\centering \includegraphics[scale=0.6]{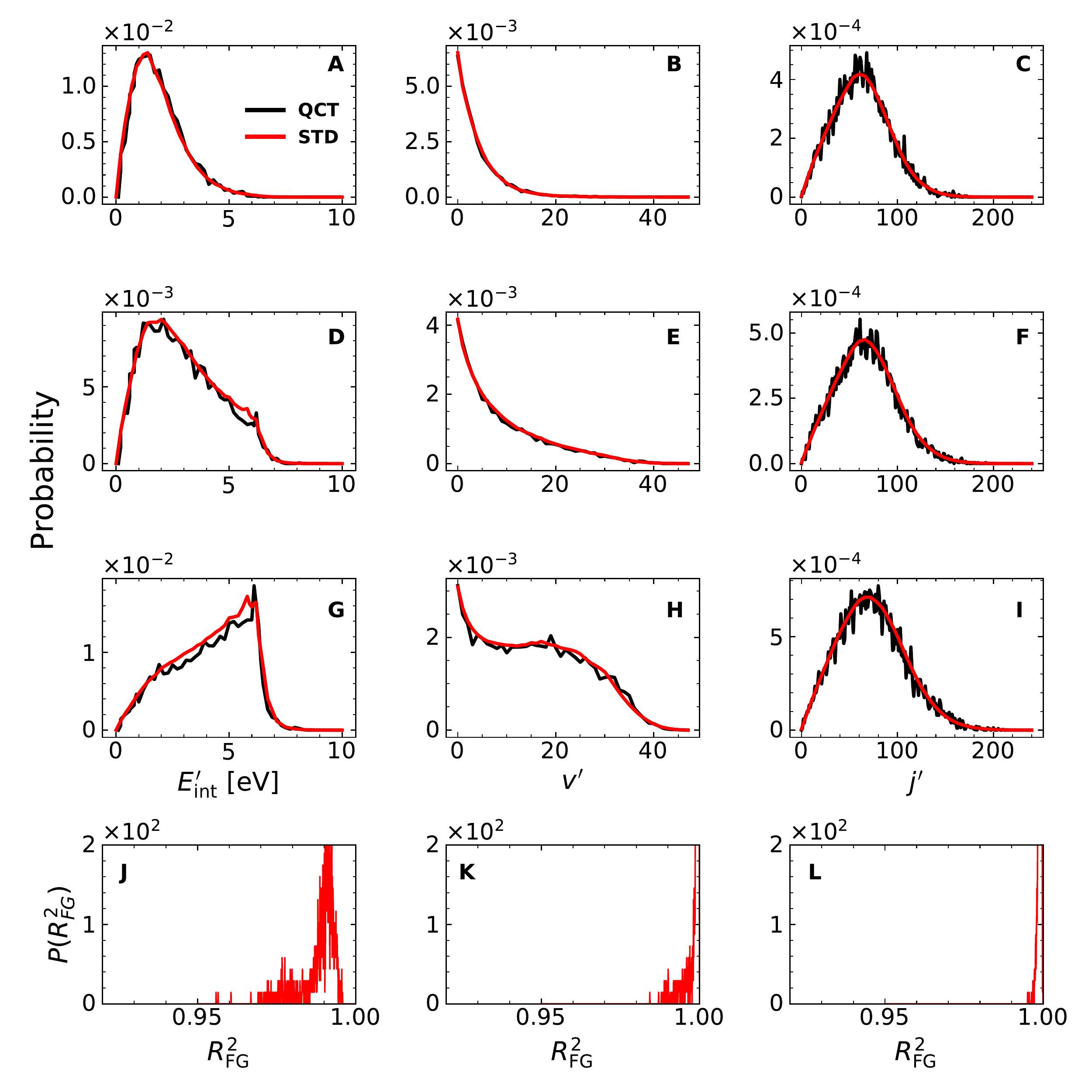}
\caption{Product state distributions for $T_{\rm trans} \neq T_{\rm
    rot}=T_{\rm vib}$ from QCT simulations compared with predictions
  from the STD model for a set of temperatures between 5000 and 20000
  K in steps of 250 K. For each set of temperatures 10000 reactant
  initial conditions are generated from Monte Carlo sampling of
  ($P(E_{\rm trans}), P(v), P(j)$). Then, final state distributions
  ($P_{i}(E_{\rm int}^{'}),P_{i}(v'),P_{i}(j')$) for each initial
  condition $i$ are obtained from evaluating the STD model and
  averaged to obtain the final state distribution ($P(E_{\rm
    int}^{'}),P(v'),P(j')$) for the particular set of
  temperatures. These distributions are then compared with the results
  from QCT simulations. Panels A-C: best performing (largest $R^2_{\rm
    FG}$) for $T_{trans}= 6500 $ K, $T_{\rm rot,vib}=5000$ K; panels
  D-F: closest to the mean of all models for $T_{trans}= 5250 $ K,
  $T_{\rm rot,vib}=9250$ K; panels G-I: worst performing (smallest
  $R^2_{\rm FG}$) for $T_{trans}= 5000 $ K, $T_{\rm rot,vib}=19000$
  K. Panels J to L report the distribution of $R^2_{\rm FG}$ values
  for the complete set $T_{\rm trans} \neq T_{\rm rot}=T_{\rm vib}$
  containing for $P(E_{\rm int}^{'})$, (K) $P(v')$ and (L) $P(j'),$
  from left to right.}
    \label{sifig:STD-DTD-set1}
 \end{figure}

\begin{figure}[h!]
  \centering \includegraphics[scale=0.6]{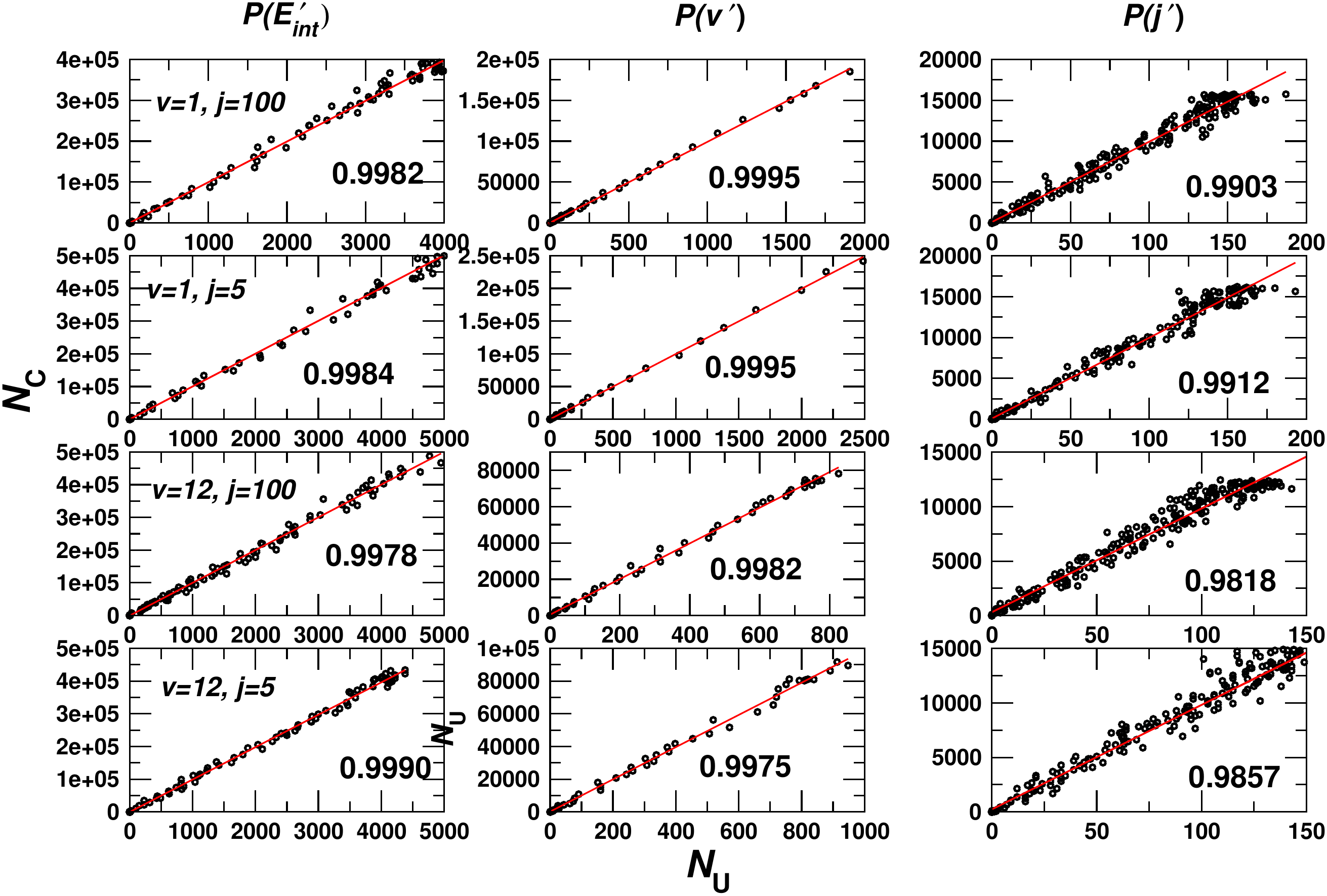}
    \caption{Correlation between converged (``ground truth'',
      $y-$axis) and unconverged ($x-$axis) QCT simulations for four
      different initial conditions with $E_{\rm trans} = 4.0$ eV as
      indicated. The ``ground truth'' is from $5 \times 10^6$
      trajectories and the unconverged data is from $5 \times 10^4$
      trajectories. The reported data compares occupation for the same
      bin for $P(E_{\rm int}')$ (left), $P(v')$ (middle), and $P(j')$
      (right). All correlation coefficients are close to 0.99 showing
      quantitative agreement.}
    \label{sifig:qctconv}
\end{figure}

\begin{figure}[h!]
  \centering \includegraphics[scale=0.6]{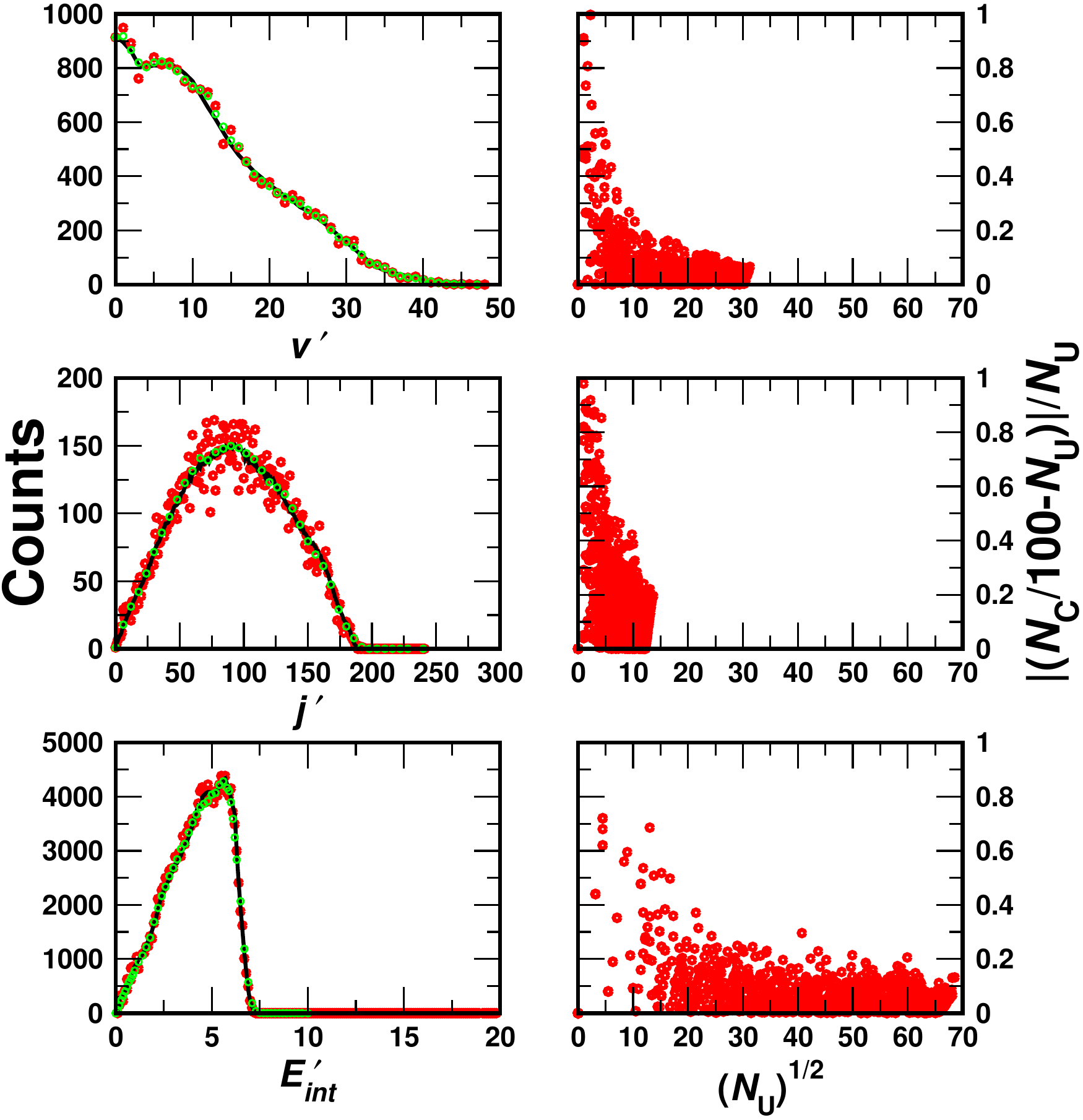}
    \caption{Product state distributions $P(v')$, $P(j')$, and
      $P(E_{\rm int}')$ for initial condition $(E_{\rm trans} = 4.0$
      eV$, v = 12, j = 5)$, from $N_{\rm U} = 5 \times 10^4$ (red,
      number of unconverged samples) and $N_{\rm C} = 5 \times 10^6$
      (black, ``ground truth'' number of converged samples)
      trajectories. The green dots are obtained by performing local
      averaging over the red data points. The reference curve obtained
      by local averaging of the unconverged QCT data matches the
      converged QCT data closely. This motivates the local averaging
      procedure performed as a data preparation step in this work. It
      allows for STD models to be trained on unconverged QCT data
      while the resulting models yield predictions that match the
      converged data closely. The right column reports the noise to
      signal ratio $\frac{(N_{\rm C}/100-N_{\rm U})/N_{\rm C}}{N_{\rm
          C}}$ of the unconverged set relative to ``ground truth'' as
      a function of $\sqrt{N_{\rm U}}$.}
    \label{sifig:qctvalid}
\end{figure}

\end{document}